\documentclass[preprint,authoryear,11pt]{elsarticle}
\usepackage[margin=0.80in]{geometry}
\usepackage{float}
\usepackage{adjustbox}
\usepackage{wrapfig}
\usepackage{graphics}
\usepackage[dvips]{color}
\usepackage{graphicx}
\usepackage{epstopdf}
\usepackage{caption}
\usepackage{subcaption}

\usepackage{amssymb,amstext,amsmath}
\usepackage{float}
\usepackage{booktabs}
\usepackage{array,ragged2e}
\usepackage[table]{xcolor}
\usepackage{multirow}
\usepackage{algorithm}
\usepackage{algpseudocode}
\usepackage{pifont}

    \makeatletter
    \def\ps@pprintTitle{%
       \let\@oddhead\@empty
       \let\@evenhead\@empty
       \def\@oddfoot{\reset@font\hfil\thepage\hfil}
       \let\@evenfoot\@oddfoot
    }
    \makeatother

\begin{document}

\begin{frontmatter}


%%\author{Name\corref{cor1}\fnref{label2}}

\title{A Simple Traffic Signal Control Using Queue Length Information}
%\maketitle
\author[rvt]{Gurcan Comert\corref{cor1}}
\ead{gurcan.comert@benedict.edu}
\author[focal]{Mecit Cetin}
\ead{mcetin@odu.edu}
\author[rvt]{Negash Begashaw}
\ead{negash.begashaw@benedict.edu}
\cortext[cor1]{Corresponding author}
%\fntext[fn1]{This is the specimen author footnote.}
%\fntext[fn2]{Another author footnote, but a little more longer.}
\address[rvt]{Department of Computer Sc., Phys., and Eng., Benedict College, 1600 Harden St., Columbia, SC 29204 USA}
\address[focal]{Department of Civil and Env. Eng., Old Dominion University, 135 Kaufman Hall, Norfolk, VA 23529 USA
\vspace{-10mm}}
\begin{abstract}
Developments in sensor technologies, especially emerging connected and autonomous vehicles, facilitate better queue length (QL) measurements on signalized intersection approaches in real time. Currently there are very limited methods that utilize QL information in real-time to enhance the performance of signalized intersections. In this paper we present  methods for QL estimation and a control algorithm that adjusts maximum green times in actuated signals at each cycle based on QLs. The proposed method is implemented at a single intersection with random and platoon arrivals, and evaluated in VISSIM (a microscopic traffic simulation environment) assuming $100 \%$ accurate cycle-by-cycle queue length information is available. To test the robustness of the method, numerical experiments are performed where traffic demand is increased and by 20\% relative to the demand levels for which signal timing parameters are optimized. Compared to the typical fully-actuated signal control, the proposed QL-based method improves  average delay, number of stops, and QL for random arrivals, by $6\%$, $9\%$, and $10\%$ respectively. In addition, the method improves average delay, number of stops, and QL by $3\%$, $3\%$, and $11\%$  respectively for platoon vehicle arrivals.
\end{abstract}

\begin{keyword}
Control, signal control, queue lengths, platoon arrivals, queue length estimation 
% MSC codes here, in the form: \MSC code \sep code
% or \MSC[2008] code \sep code (2000 is the default)

\end{keyword}

\end{frontmatter}

% \linenumbers
%\begin{linenumbers}
\section{Introduction}
\label{sctintro}
%\lable{sctIntro}
In real-world traffic operations, demand at each intersection approach is subject to significant fluctuations throughout the day. To adapt these changes, phase lengths need to be adjusted in real-time to minimize delays and maximize throughput. In order to notify the traffic signal controller that a vehicle is calling a green phase (during a red interval) and extend the green time for an approaching vehicle (during a green interval), the common practice is to install fixed detectors (i.e., inductive loops or video cameras) to detect vehicle presence at the stop-bar. The use of vehicle detection to call and extend phases is commonly referred to as actuated control. A typical actuated control framework operates within the constraints of minimum (min) and maximum (max) green times for each phase. If the max green times are not high enough to accommodate the demand for a phase then residual queues can occur during peak traffic flow.  It can sometimes take numerous cycles to fully disperse the residual queue. In this paper we present a method to minimize the average delay at an intersection. The method uses queue length (QL) measured in real time on each approach to update the duration of the max green interval.
%However, residual queues can occur during peak traffic flow if the max green times are not high enough to accommodate the demand %for a phase whenever the entire queue is not discharged within the max green time and the signal maxes out. It can sometimes take %numerous cycles to fully disperse the residual queue, usually when only the upstream demand decreases. 
%This study presents a method to adapt the duration of the max green interval with queue length (QL) measured in real-time on each %approach to minimize the average intersection delay. 

Actuated signal control operates based on a simple binary input from vehicle detectors (vehicle presence/non-presence). In the paper by Lin ( \cite{Lin1985}), green phases are extended only based on the detector inputs serving these phases (actuation) and conflicting calls on the competing phases. There is no distinction between one vehicle  versus tens of vehicles on the conflicting movement. 

%For example, green phases are extended only based on the detector inputs serving these phases (actuations) and conflicting calls on %the competing phases (\cite{Lin1985}).  There is no distinction between one vehicle on the conflicting movement versus tens of %vehicles.  
An adaptive max green feature is available in some manufacturer's traffic signal controllers. According to National Electronic Manufacture Association 1202 (\cite{NTCIP2004}), definition for actuated traffic signal control is the adaptive green operation that is defined as a cycle-by-cycle max green interval adjustment within an upper and lower limit. This feature allows the max green time for a phase to be increased by a set time once a phase maxes out for a defined number of consecutive cycles. In a previous work, \cite{Yun2007} has shown that this method has the benefit in accommodating peak traffic. The method is somewhat slow to respond to the field conditions due to its fixed operational constraints (e.g., the adaptive max parameters do not change) (\cite{Engelbrecht2003}). The proposed method in this research directly incorporates QL information, rather than using "max out" occurrences as a surrogate for the QL information.

%Although this method has provided some benefit in accommodating peak traffic as shown in previous research (\cite{Yun2007, %Engelbrecht2003}), it is somewhat slow to respond to the field conditions due to  its fixed operational constraints (e.g., the adaptive %max parameters do not change). 
 
Currently, there are several surveillance technologies like loop detectors, video cameras, wireless vehicle detectors, and probe (or connected) vehicles.  Accuracy of such detectors changes under different conditions. Information obtained from fixed vehicle detection is limited to the physical location where it is  implemented along the approach.  Fixed detectors are capable of measuring QLs through numerous inductive loops (or multiple video detection zones) that are implemented from the stop-bar to a sufficiently long distance upstream of the intersection. From loop detectors, queue length estimators can be achieved within $15\%$ in mean absolute percent errors (\cite{liu2009real}). This is not feasible with inductive loops due to the cost associated with the installation. It is also difficult to measure queue lengths with video detection due to the diminishing field of view, unless multiple video cameras are installed at different upstream locations.  

Newer technologies allow the estimation of QLs in nearly real-time. Recent developments in vehicle detection technologies provide the ability to measure QLs more accurately at signalized intersections (\cite{Lin1985,Travis}). Vehicles equipped with wireless communications (e.g., DSRC or upcoming 5G), in particularly connected vehicles (CV), and some radar detection devices can provide this capability (\cite{smartsensor}). The benefit of CVs over loop detectors and video cameras at signalized intersections was studied by Smith et.al. (\cite{smith2010intellidrivesm}). After $25\%$ market penetration rate, connected vehicles showed $2.5$ and $1.1$ benefit to cost ratio over loop detectors and video cameras, respectively. At $100\%$, CV percent improvement in delays was found to be $6.8\%$ (\cite{smith2010intellidrivesm}).

%which provides the motivation for the proposed methodology.
 
%Recent developments in vehicle detection technologies provided the ability to measure QLs more accurately at signalized intersections %(\cite{Lin1985,travis}).

There are numerous research efforts to develop methods to optimize traffic signals (\cite{Park2002, Liu2002, Cassidy1998, Abbas2006, Mirchandani2001, Bullock2004, Comert2009TRB}). However, there is still no accepted methodology for optimizing or enhancing the performance of a signalized intersection using  QL data. QL is usually utilized to measure the performance of signalized intersections. It is used to estimate delays and travel times (\cite{Gartner2002,Mirchandani2007,lee2015real,feng2015real}). It is shown in an eco-signaling application, queue lengths are utilized for speed recommendation which are high negatively correlated (\cite{yang2017}). With known queue lengths,  adaptive signal control was also studied in a reinforcement learning scheme (\cite{genders2016using,gao2017adaptive}). If true QLs can be obtained as feedback, \cite{comert2019grey} showed that QLs for next time interval can be predicted within $5$ meters ($m$) accuracy. Assuming Poisson arrivals, Zhang and Wang (\cite{Zhang2011}) developed a method  to  optimize min green and max green parameters of an actuated control using real-time QL information (i.e., utilizing the information in the constraint of the optimization).  In our approach,  we do not assume Poisson arrivals. Partial information about vehicle arrivals can also be found  in queue length estimation.   

%The only similar approach to this paper is the method by 
Past research also gives detailed insights about the distributions of CV information types that can be utilized for more complex models (e.g., multilane intersections) and estimating delays for signal control parameters (\cite{comert2013simple, tiaprasertqueue}). Similar results were used as input for QLs/delay approximations in new signal control strategies (\cite{Comert2009TRB,smith2010intellidrivesm,goodall2013traffic}). Accuracy of QL estimation for the traffic signal control under connected vehicle (CV) framework was also investigated by Tiaprasert et. al (\cite{tiaprasertqueue}). Allocating discrete wavelet transform, the authors obtained enough accuracy at $30\%$ CV penetration rate. Similar market penetration rate was observed in (\cite{comert2013effect,comert2016queue}).

Connected signal control is one of the priority applications as $80\%$ (250,000) of the Nation's signal network and $90\%$ of the vehicles are projected to be equipped by 2040 (\cite{wright2014national}). Accurate, real-time, localized, traveler information will be available on at least $90\%$  of roadways. Next-generation, multimodal, information-driven, active traffic management will be deployed system-wide. Thus, step-by-step inference based (up to 20\% of signals to be equipped by 2025) and fully information based (with some filtering) need to be developed. Overall, it is necessary to develop signal control algorithms that focus on various points of connected systems which vary from communication protocols, use of effective and different data types, to security and reliability of such components (\cite{dey2016vehicle}). Adaptive actuated traffic signal control was introduced by changing unit extensions and max greens using arrival profiles. Zhang and Recker  (\cite{zheng2013adaptive}) showed $17.3\%$ and $16.2\%$ improvements in max QL and delay respectively. In another study under CV framework, Feng et al. (\cite{feng2015real})  proposed an algorithm that outperforms the typical actuated control when the penetration rate is more than $50\%$. When all vehicles are connected, the total delay is improved by $12\%$ on average when minimizing total vehicle delay and $11\%$ when minimizing QL. In our proposed simpler method, no cycle-by-cycle optimization is required. 
\par
Estimation of queue length problem was discussed in the context of performance rather than cycle-by-cycle control and improvement of signal timing via weighing queues at competing approaches relatively. In this paper, we aim to fill the gap of simple adaptations of signal timing to demand changes shown in queue lengths. Hence, we propose a queue length-based adaptive signal control. The proposed method is tested in VISSIM, a state-of-the-art microscopic traffic simulation environment (\cite{VISSIM2012}), for an isolated intersection. This platform is a critical and valuable tool in assessing the benefits of the queue data on signal operations.  It allows testing different scenarios under different assumptions while representing vehicular movements realistically. The proposed  method is tested with both random arrivals and platoon vehicle arrivals under various demand scenarios. Platoon arrivals are generated with a signalized intersection located $650$ $m$ upstream of the intersection where queue lengths are measured.   
\vspace{-10pt}
\section{Queue Length-based Signal Control}
\label{sctprb}
In the proposed method,  the maximum (max) green times in each cycle are calculated based on a simple formula as described  in equation (\ref{eqn_gmax1}) below. In a typical actuated signal control, there are three key parameters for each phase: minimum (min) green, maximum (max) green, and vehicle (unit) extension. For given input volumes,  these three parameters are set to fixed values based on the output from a signal optimization program (e.g., Synchro). The max green timer counts down when there is a call on the conflicting movements.  If there is no vehicle on the conflicting phases for some duration then the displayed green time can be longer than the preset green times

When a phase is extended up to the max green time because of high demand,  the phase is said to be  \textit{maxed out}. If one of the phases maxes out many times and the other phases do not then this may be an indication of inefficient allocation of the capacity. A potential solution is to make max green adaptive to the prevailing traffic conditions. 
Yun et al. (\cite{Yun2007}) provide an adaptive control feature where the max green is adjusted based on the number of max out events that occured in the past few cycles. 

%describe the adaptive maximum feature in some modern signal controllers that allows the max green to be adjusted depending on the %number of max out events that has taken place in the past few cycles. 

Alternatively, the max green time for each phase can also be determined adaptively  based on QL data.  One approach is to set the max green for each phase based on the QL observed at the end of red interval that immediately proceeds the green time. For a simple intersection with two one-way streets, the max green for the major and minor streets can be expressed as  a function of the QL as follows: 

\begin{equation}
G_{max_1,k+1}= \max\left(LB_1,\quad\min\left[UB,\quad\beta\left(\frac{N^{2}_{1,k+1}}{(N_{1,k+1}+N_{2,k}}\right)\right]\right)
%G_{max_1,k+1}= \max\left(LB_1, \min[UB,\beta(N^{2}_{1,k+1}/(N_{1,k+1}+N_{2,k}))]\right)				       
\label{eqn_gmax1}  
\end{equation}
\begin{equation}
G_{max_2,k+1}= \max\left(LB_2,\quad\min\left[UB, \quad\beta\left(\frac{N^{2}_{2,k+1}}{(N_{2,k+1}+N_{1,k}}\right)\right]\right)
%G_{max_2,k+1}=\max\left(LB_2, \min[UB,\beta(N^{2}_{2,k+1}/(N_{2,k+1}+N_{1,k}))]\right) 				       
\label{eqn_gmax2}  
\end{equation}
where, 
\newline
$G_{max_1,k+1}$ is the max green time of major street at cycle $k+1$,\\
$G_{max_2, k+1}$ is the max green time of minor street at cycle $k+1$,\\
$N_{1,k+1}$ is the queue length (in number of vehicles) on major street at cycle $k+1$,\\
$N_{2,k+1}$ is the queue length (in number of vehicles) on minor street at cycle $k+1$,\\
$N_{1,k}$ is the queue length (in number of vehicles) on major street at cycle $k$,\\
$N_{2,k}$ is the queue length (in number of vehicles) on minor street at cycle $k$,\\
$\beta$ is a model parameter, \\
$LB_1$ is lower bound on green time for major street,\\
$LB_2$ is lower bound on green time for minor street,\\
$UB$ is upper bounds for max green (the same for major and minor streets)\\

%In Eq. (\ref{eqn_gmax1}), the expression in the square brackets being multiplied by the parameter $\beta$, is the product of the QL of the current phase with the proportion of the QL of the current phase relative to the total of the QL of the current phase and the QL of the opposing traffic. Similarly, in Eq.~(\ref{eqn_gmax2}), the expression in the square brackets, multiplying the parameter $\beta$, is the product of the QL of the current phase with the proportion of the QL of the current phase relative the total of the QL of the current phase and the QL of the opposing traffic.

In equations (\ref{eqn_gmax1}) and (\ref{eqn_gmax2}),  if there are more than two phases, QLs on other approaches can be included in the denominators of the fractional expressions. QLs of the conflicting movements in cycle $k+1$ are approximated (predicted) by the QLs observed in the previous cycle. Therefore, there is an element of randomness in this method.  

%It should be noted that when calculating the max green for the next phase in cycle $k+1$ , the QLs on conflicting movements are past the observations in cycle $k$ which may not be identical to the queues to be observed in cycle $k+1$ on these conflicting phases. In other words, 

Eqs. (\ref{eqn_gmax1}) and (\ref{eqn_gmax2}) have three key parameters to be optimized: $LB_1$, $LB_2$, and $\beta$. $UB$ is set to a large value that represents the maximum allowable time (e.g., $5$ minutes in the experiments) that the signal phase can stay green under extreme demand scenarios. $LB_1$, $LB_2$, and $\beta$ are optimized offline for some given intersection demand scenarios (\cite{comert2009incorporating}). For completeness, the same random seed is used in simulating both methods in VISSIM to ensure that the generated vehicles have the same headways. 

%Grid search results for the parameters of the typical actuated control and QL-based actuated control, respectively, for the demand profile 1. The optimum parameters are found based on the average intersection delay (seconds per vehicle). Optimum parameter values for all three base volumes are shown later in Table~\ref{tab_par}. 

The proposed model control parameters are determined after running numerous simulations in VISSIM with uniform increments. The set of parameters that yield the lowest average delay are given in Table~\ref{tab_par}. The optimal results of the QL-based method shown in the table provide improvements in average delays over the typical actuated control. It is found that the value of the parameter $\beta$ is the same in all three cases. Other $\beta$ values were tried and there were no improvements. 
%For example, neither $beta=3$ nor $\beta=5$ provide better delay values at base volumes.
 More demand profiles need to be tested  before one can make any generalizations about $\beta$.

\subsection{Selection of the Parameters in Equations \ref{eqn_gmax1} and \ref{eqn_gmax2} }
\label{sct_parsel}
\begin{figure}[ht!]
\centering
\includegraphics[scale=.50]{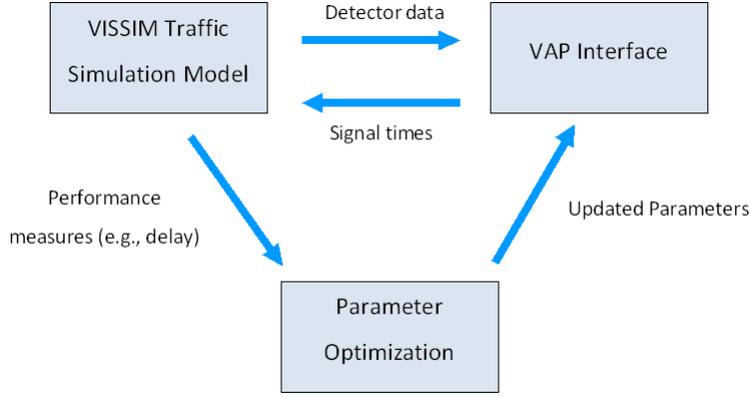}
\caption{Parameter optimization framework}
\label{fig_frame}       
\end{figure}

To choose the parameters in Eqs. (\ref{eqn_gmax1}) and (\ref{eqn_gmax2}), three main tools are used: a microscopic traffic simulation platform to generate measures of performance (e.g., delay, number of stops, and QLs), an interface to change traffic signal times, and an optimization algorithm to determine the best parameters. These three components and the flow of information  between them are shown in Fig. \ref{fig_frame}. VISSIM is the main tool that simulates the movements of all vehicles at the intersection. Vehicle Actuated Programming (VAP)  interface is a component within VISSIM that allows the user to manipulate the signal times based on data from  the detector. The QLs (i.e., $N_{1,k+1}$, $N_{2,k+1}$) are measured in VISSIM on each approach. Enough node evaluation space is covered such that for any demand scenario the QLs never extend beyond the evaluation area. The average intersection delay calculated by VISSIM is used as the performance measure to optimize the parameters. Since delay does not have a closed form solution, in order to ensure that a good solution is found, a grid search is used where the parameter values are changed by fixed increments. The parameters that give min delay values were selected for both control methods.
\vspace{-15pt}
\subsection{Queue Length Estimation}
\label{QLE}
In the proposed method, QLs, $N_{1,k}$, $N_{1,k+1}$, $N_{2,k}$, and $N_{2,k+1}$ are the dynamical inputs. In case, connected vehicles are utilized for QL information, it is important check at what market penetration level, adequate accuracy can be achieved for given estimators. Control results are given for known or true QL information, that is, for best case scenario how much can be saved with proposed control. Errors of the adopted estimators are calculated using VISSIM simulations. Queue length estimation with known and unknown hyper parameters (i.e., arrival rate $\lambda$ and probe percentage $p$) are given in Table~\ref{tab_est}. These estimators are the best performing combinations obtained from \cite{comert2016queue}. Estimation errors obtained from simulations are presented in tables and figures below.

\begin{table}[h!]
\centering
\caption{Estimators for $p$ and $\lambda$}
\label{tab_est} 
\scalebox{0.80}{    
\begin{tabular}{l c c c}
%p{1.5cm}p{2cm}p{2cm}p{2.4cm}p{2cm}
\hline\noalign{\smallskip}
Information & $\hat{p}$ & Information & $\hat{\lambda}$ \\
\noalign{\smallskip}\hline\noalign{\smallskip}
$M,L$ & $\hat{p}_{1}=m/l$ & $L$ & $\hat{\lambda}_{1}=l/R$ \\
$M,L,T$ & $\hat{p}_{2}=mt/(mt+(l-m)R)$ &$L,T,M$ & $\hat{\lambda}_{2}=(l-m)/t+m/R$ \\
\noalign{\smallskip}\hline\noalign{\smallskip}
\end{tabular}}
\end{table}
\subsubsection{No Overflow Queue Case: $Q=0$}
\label{sctqlewoq}
Estimator of the total QL at the end of red duration given location ($L=l$), queue joining time of the last CV ($T=t$), and number of CVs in the queue ($M=m$) is written as sum of two random variables $N'$ and $N''$. Random variable $N'$ is the queue up to the last CV and $N''$ is the queue after the last probe vehicle (Eq.~(\ref{qle})). When time index $i$ is the cycle number then the estimator is cycle-to-cycle queue estimator. For an alternative time interval, scanned $(L,T,M)$ can be used for estimation. Certainly, this is a lower bound as some probes may have already left the intersection. Incorporating the counted discharged vehicles, the problem can be alleviated (\cite{hao2014cycle}). For implementation, QL estimated at the end of red can be used for timing the green duration for the signal and a lower bound to the average QL for broader signal performance measures. Under the Poisson assumption with known parameters, estimator can be expressed as in Eq.~(\ref{epqle}). Note that for a multilane formulation, the following estimators can be used for signal timing as max QLs for each lane.
\begin{equation}
E(N|L=l,T=t,M=m,Q_{i}=0)=E(N'|L=l,T=t,M=m)+E(N''|L=l,T=t,M=m)
\label{qle} 
\end{equation}
\begin{equation}
E(N|L=l,T=t,M=m,Q_{i}=0)=l+(1-p)\lambda(R-t)
\label{epqle}
\end{equation}

Without overflow queue, the total QL with unknown arrival rate and probe proportion can be estimated using Eq.~(\ref{eqn_qle}). Pairs of the estimators that are considered in the  numerical examples are given in Table~\ref{tab_eQLE}.
\begin{equation}
E(N|l,t,m,Q_{i}=0)=l+(1-\hat{p})\hat{\lambda} (R-t)
\label{eqn_qle}
\end{equation}

\begin{table}[h!]
\centering
\caption{Estimators in queue length estimation}
\label{tab_eQLE}
\scalebox{0.8}{
\begin{tabular}{l c c}
\hline\noalign{\smallskip}
Estimator &Combinations & $E(N|l,t,m,Q_{i}=0$\\
\noalign{\smallskip}\hline\noalign{\smallskip}
Est. 1&$p,\lambda$& $l+\lambda(1-p)(R-t)$\\
Est. 2&$\hat{p}_{1},\hat{\lambda}_{1}$& $l+(l-m)(1-t/R)$\\
Est. 3&$\hat{p}_{2},\hat{\lambda}_{2}$& $l+(1-\frac{mt}{mt+(l-m)R})((l-m)/t+m/R)(R-t)$\\ 
\noalign{\smallskip}\hline\noalign{\smallskip}
\end{tabular}}
\end{table}

Table~\ref{tab_woQ} shows the QL estimation errors obtained from VISSIM evaluations for demand levels of $600$, $700$, $800$, $900$, and $985$ vph. In the table, root mean squared errors (RMSE) and $\%RMSE/Est_i$ are also given. Average QL will depend on VISSIM queue definition. In the numerical experiment, default \textit{in queue}=$I(speed\leq10$ kilometers per hour) was adopted. From the table, at 500 vehicles per hour, $6.25$ vehicles per 45 seconds red duration is expected. Similarly for others $7.5, 8.75, 10.00, 11.25$, and $12.31$ vehicles per red duration were expected. Overflow queue becomes more important when $\rho>0.80$. On the average, we observe queue length of $1.25$ for $\rho=0.88$ and queue length of $5.70$ for $rho=0.98$. Accuracy of the estimators change from $\pm1.5$ to $\pm3.45$ vehicles after $\rho>0.88$. Overall, estimator 2  provide accuracy between $12\%$ to $19\%$ of the true average QLs.         
 
\begin{figure}[h!]
\centering
\begin{subfigure}{.49\textwidth}
 \centering
\includegraphics[width=1\linewidth]{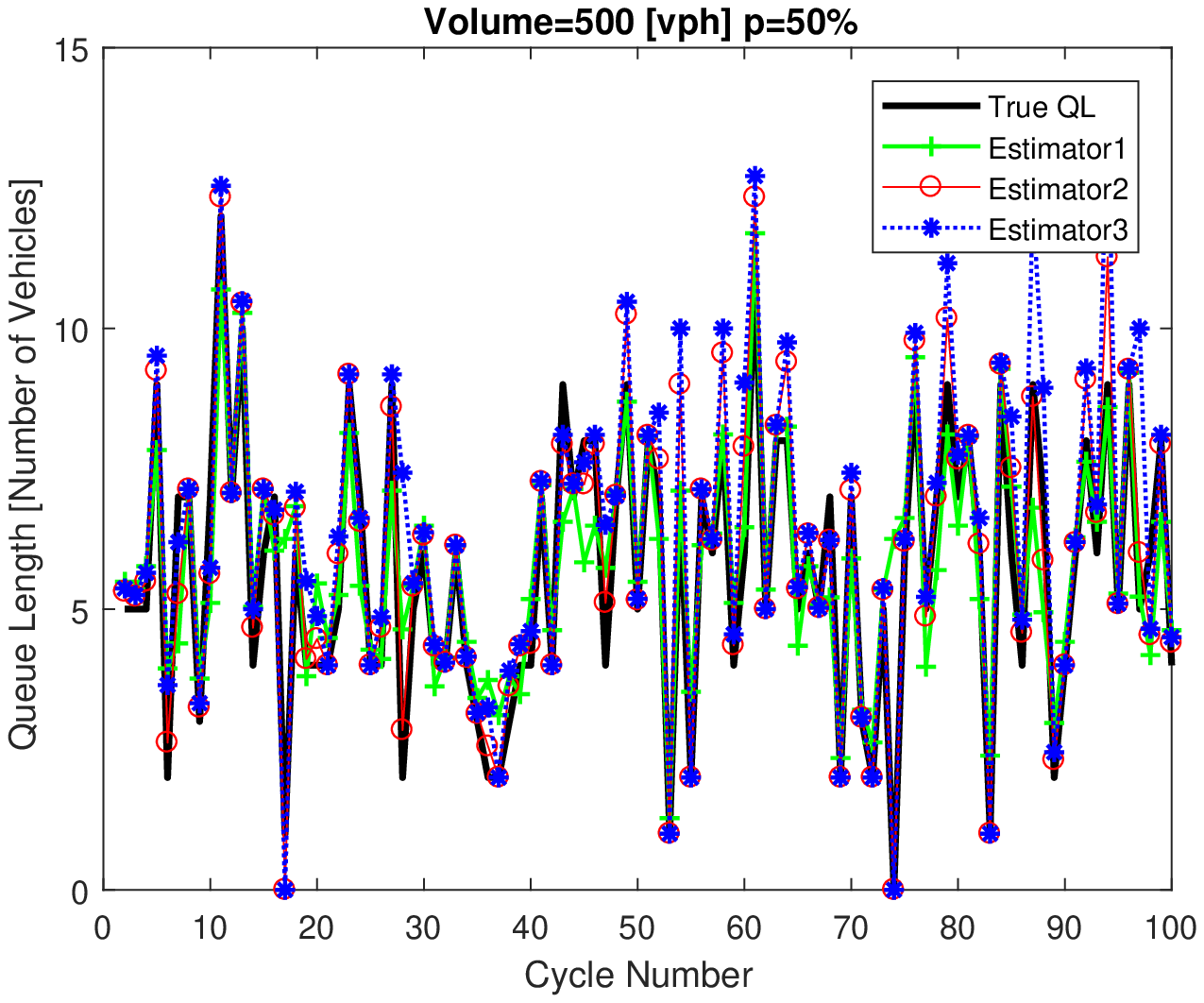}
  \caption{Estimation at $\rho$=0.50}
  \label{fig_5}
\end{subfigure}
\begin{subfigure}{.49\textwidth}
\centering
  \includegraphics[width=1\linewidth]{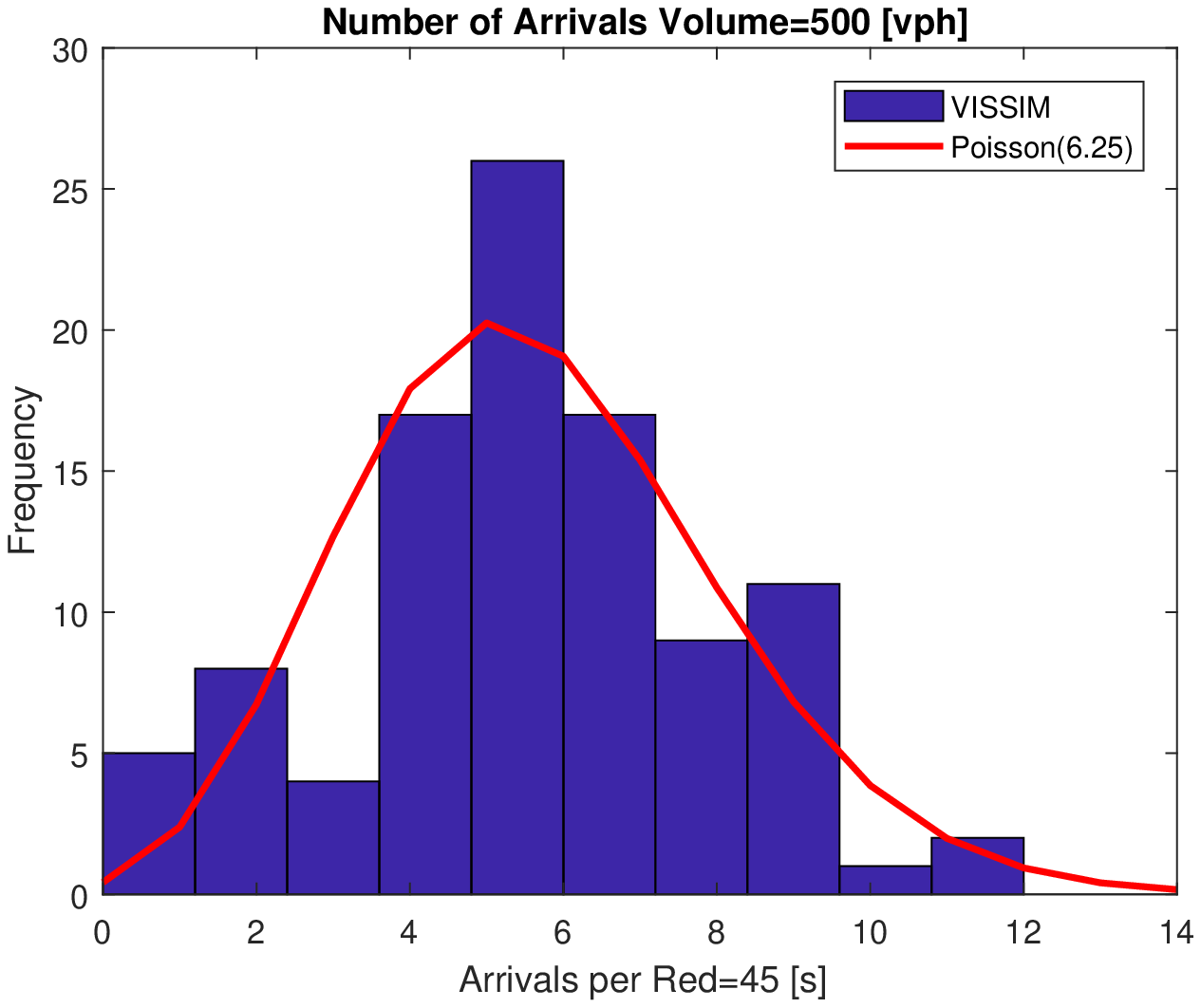}
\caption{Distribution at $\rho$=0.50}
  \label{fig_5h}
\end{subfigure}%
\\
\begin{subfigure}{.49\textwidth}
 \centering
\includegraphics[width=1\linewidth]{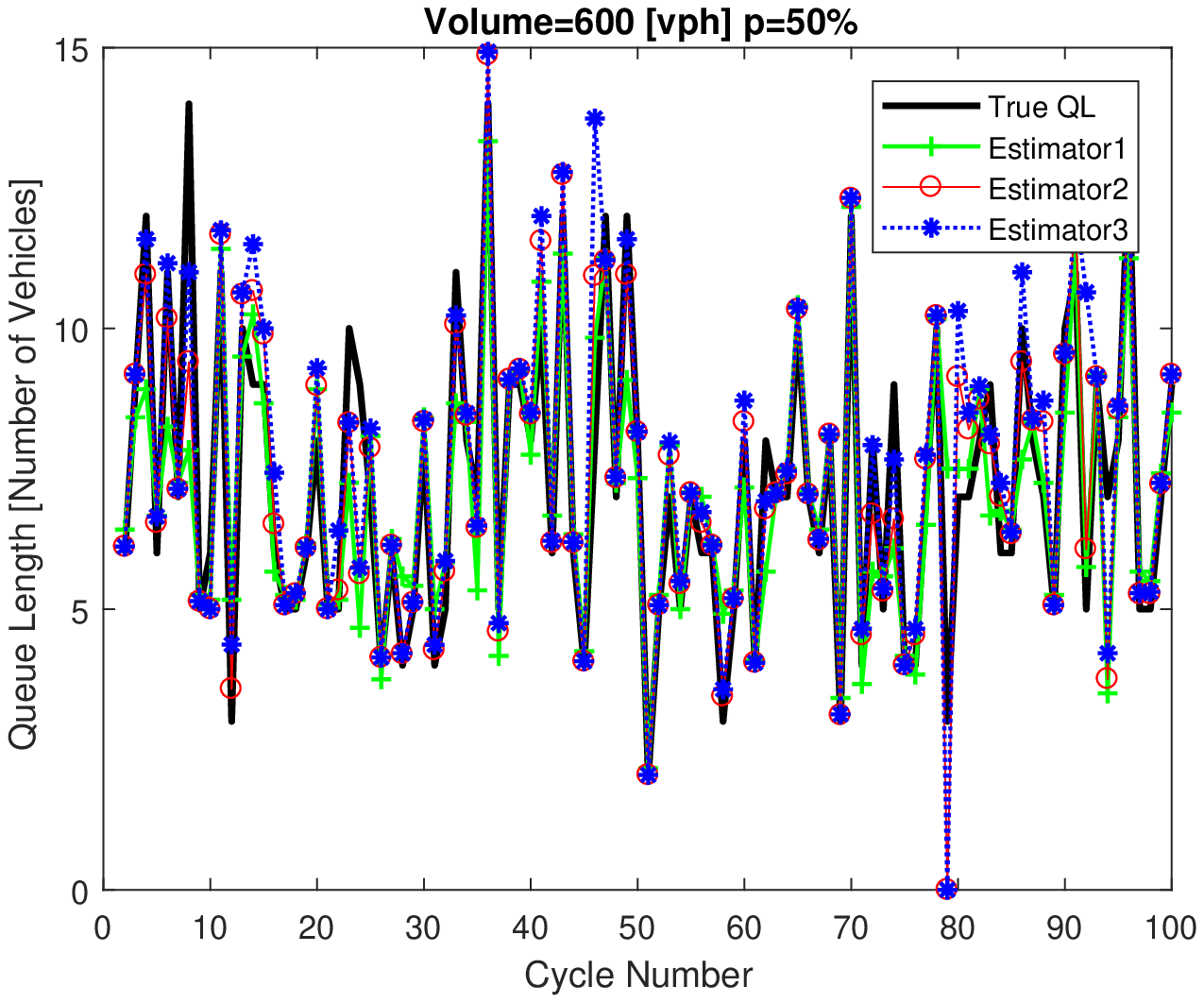}
  \caption{Estimation at $\rho$=0.60}
  \label{fig_6}
\end{subfigure}
\begin{subfigure}{.49\textwidth}
 \centering
\includegraphics[width=1\linewidth]{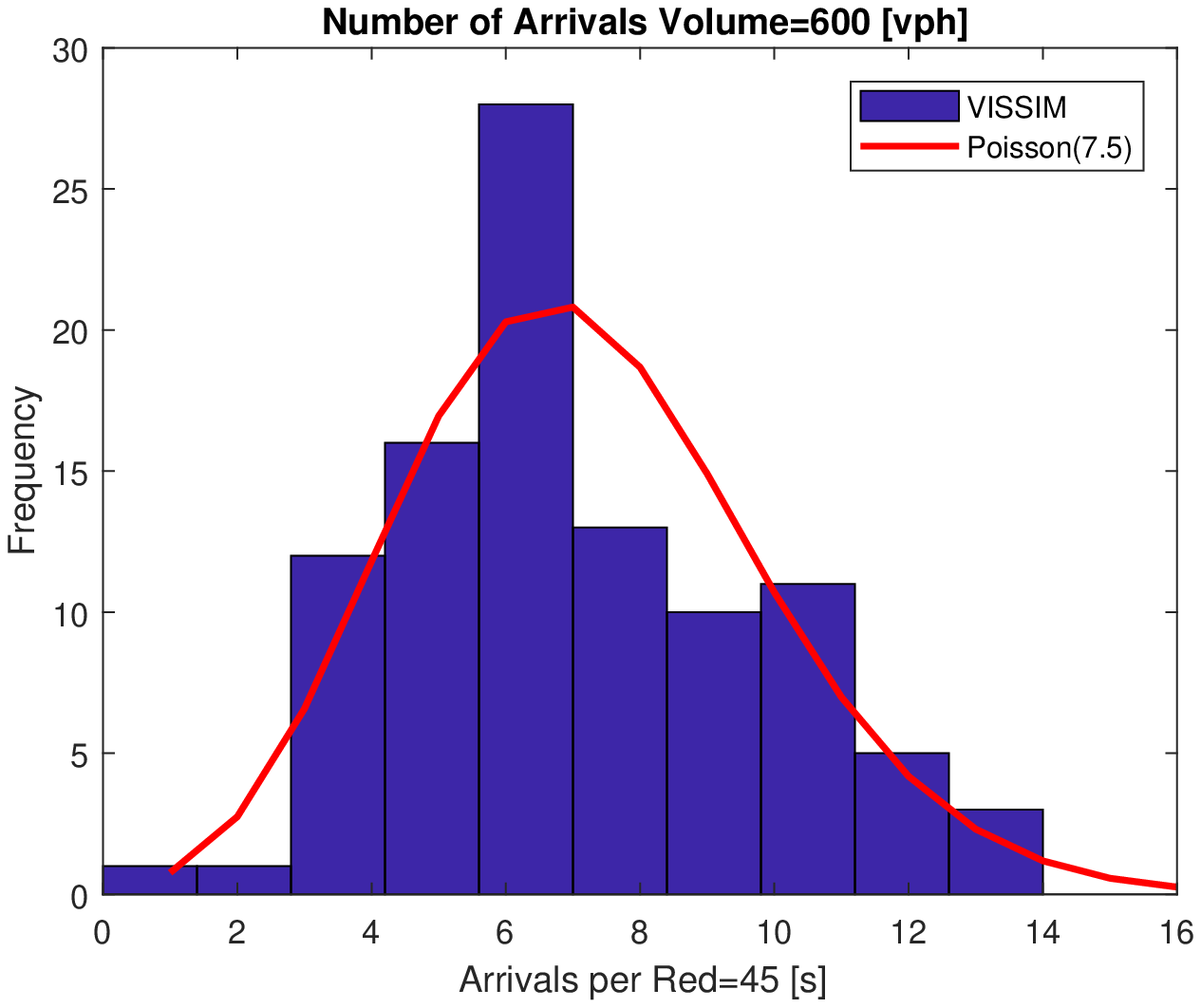}
  \caption{Distribution at $\rho$=0.60}
  \label{fig_6h}
\end{subfigure}
\\
\begin{subfigure}{.49\textwidth}
 \centering
\includegraphics[width=1\linewidth]{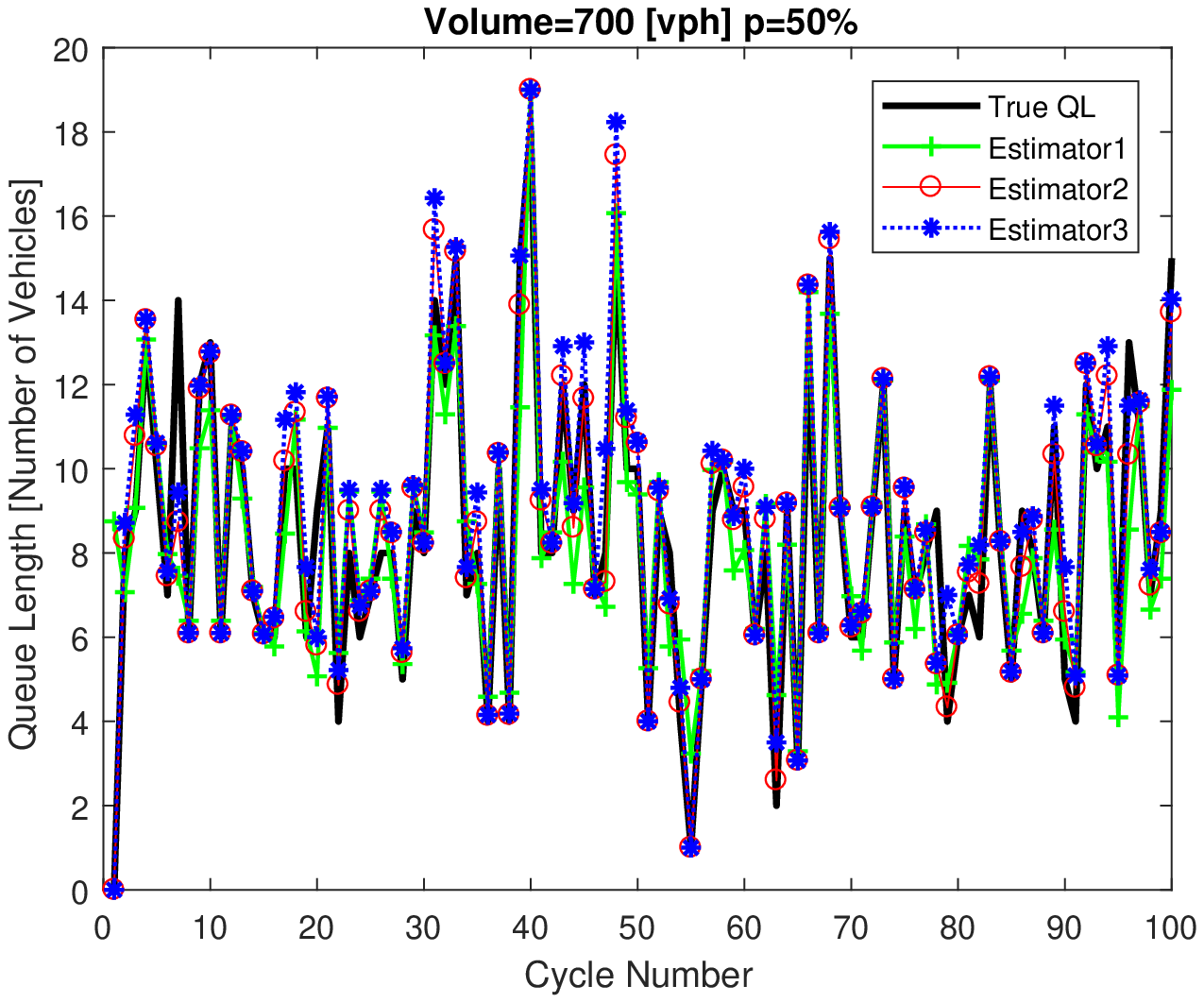}
  \caption{Estimation at $\rho$=0.70}
  \label{fig_7}
\end{subfigure}
\begin{subfigure}{.49\textwidth}
 \centering
\includegraphics[width=1\linewidth]{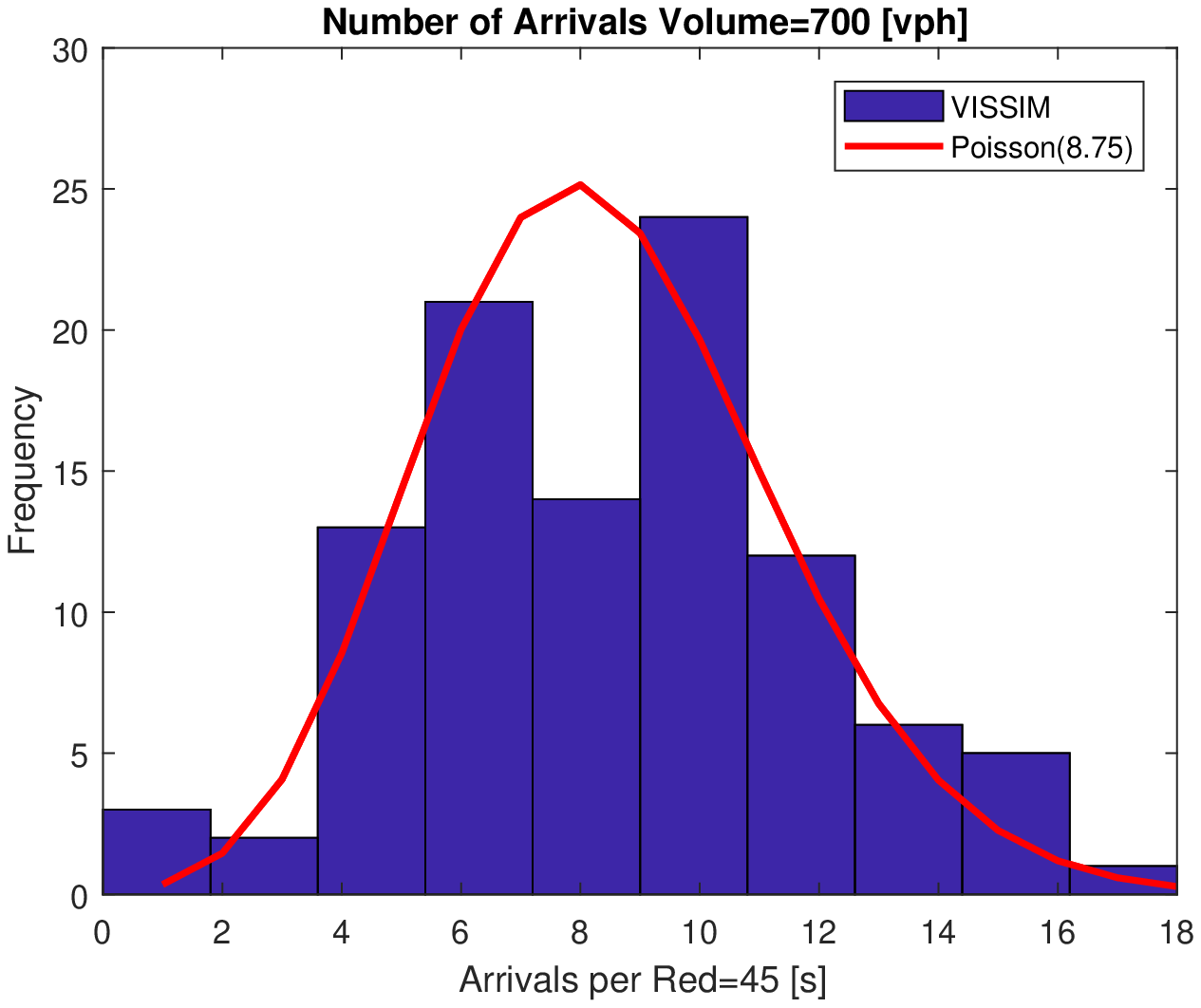}
  \caption{Distribution at $\rho$=0.70}
  \label{fig_7h}
\end{subfigure}
\caption{Cycle-by-cycle QL estimation for different $\rho<0.80$ values at $p=50 \%$}
\label{fig_low}
\end{figure}

\begin{figure}[h!]
\centering
\begin{subfigure}{.49\textwidth}
 \centering
\includegraphics[width=1\linewidth]{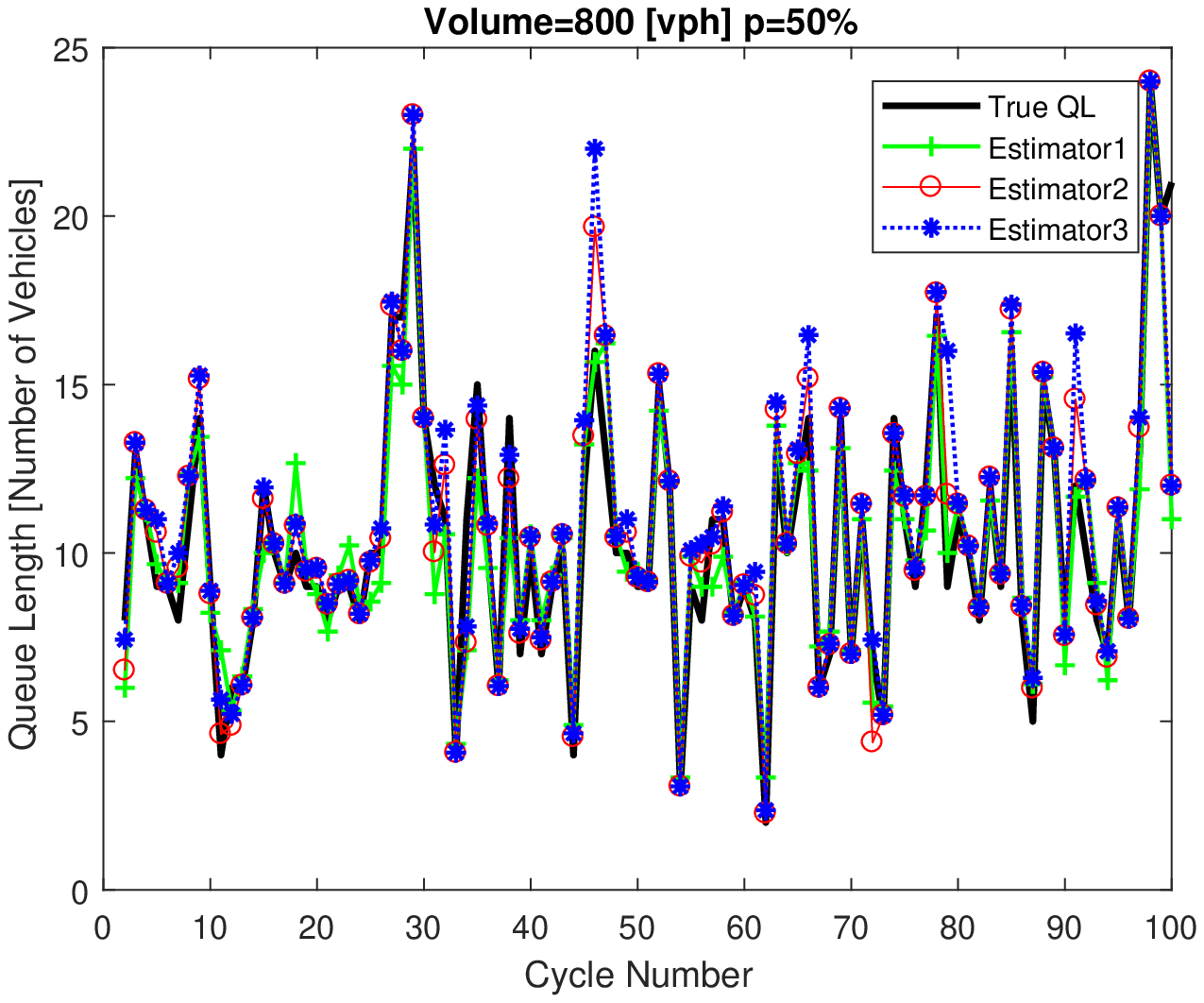}
  \caption{Estimation at $\rho$=0.80}
  \label{fig_8}
\end{subfigure}
\begin{subfigure}{.49\textwidth}
\centering
  \includegraphics[width=1\linewidth]{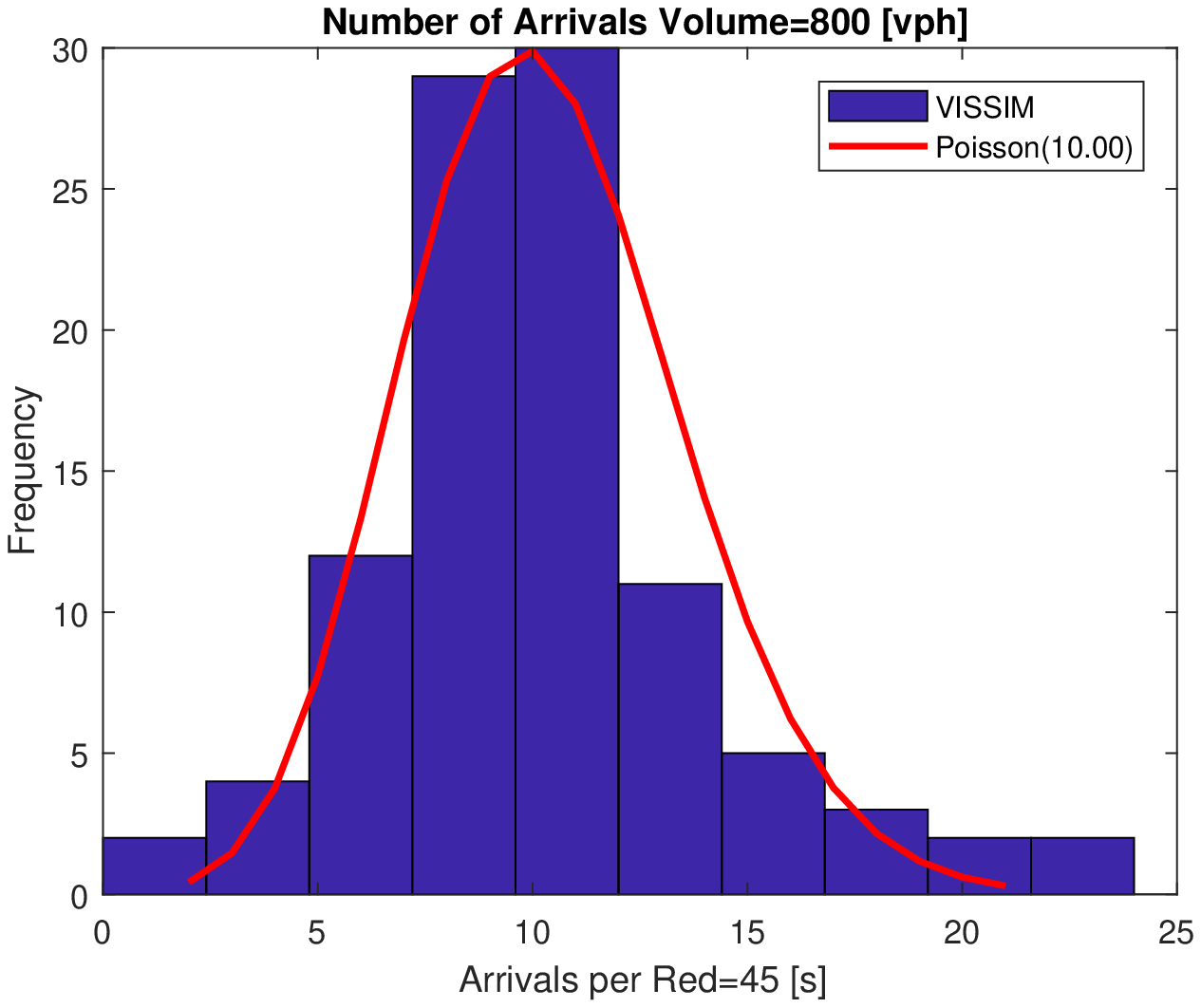}
\caption{Distribution at $\rho$=0.80}
  \label{fig_8h}
\end{subfigure}%
\\
\begin{subfigure}{.49\textwidth}
 \centering
\includegraphics[width=1\linewidth]{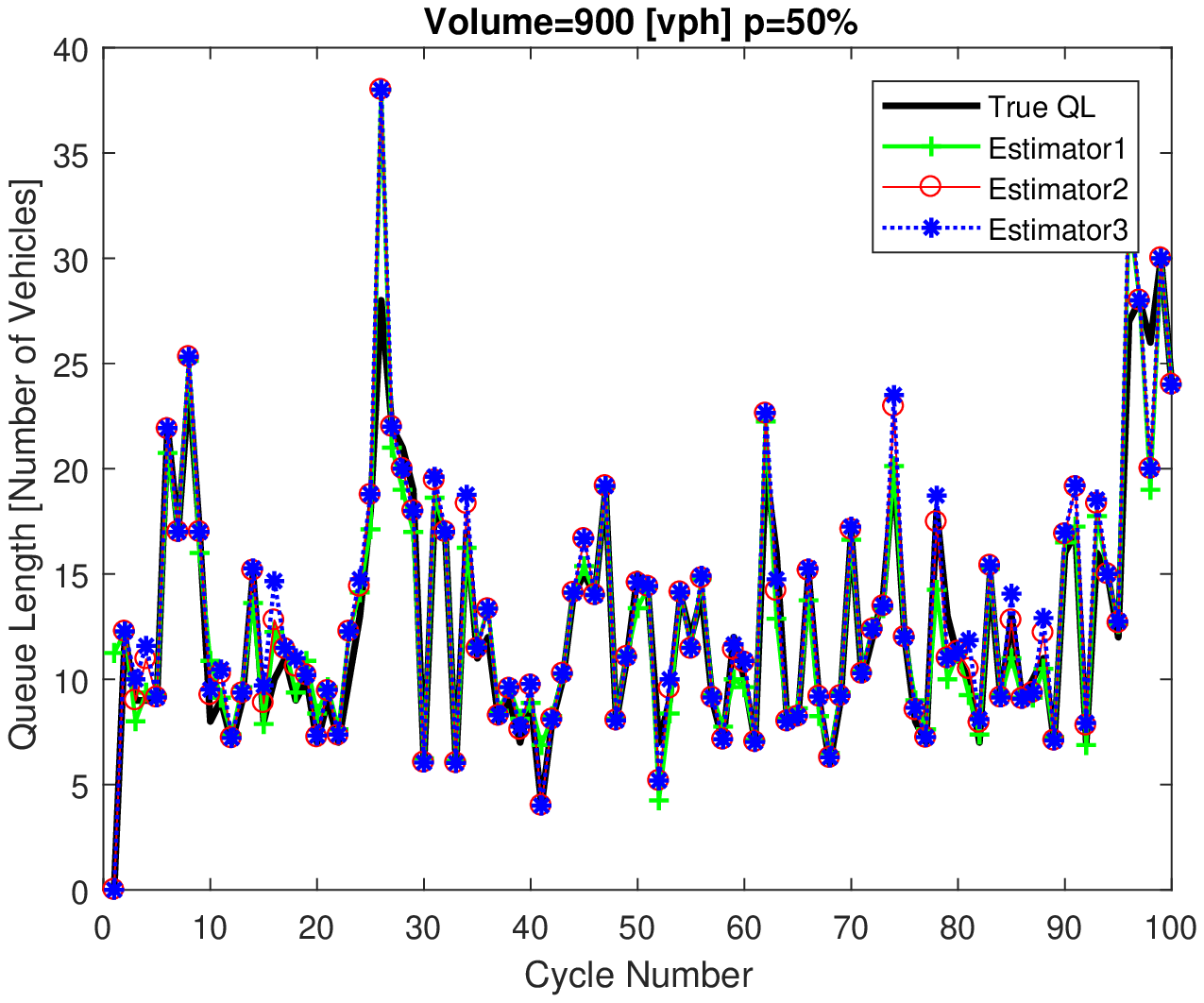}
  \caption{Estimation at $\rho$=0.88}
  \label{fig_9}
\end{subfigure}
\begin{subfigure}{.49\textwidth}
 \centering
\includegraphics[width=1\linewidth]{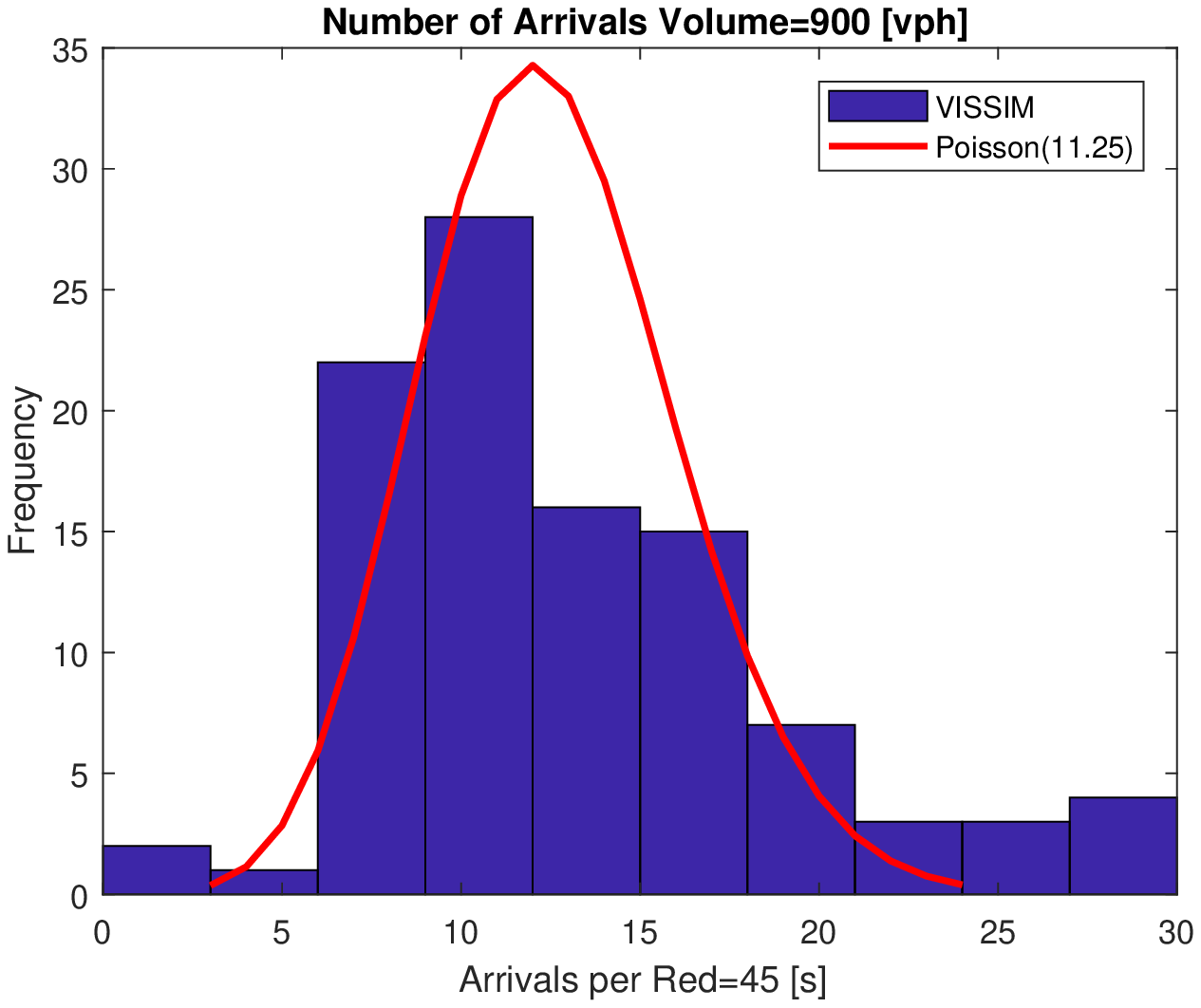}
  \caption{Distribution at $\rho$=0.88}
  \label{fig_9h}
\end{subfigure}
\\
\begin{subfigure}{.49\textwidth}
 \centering
\includegraphics[width=1\linewidth]{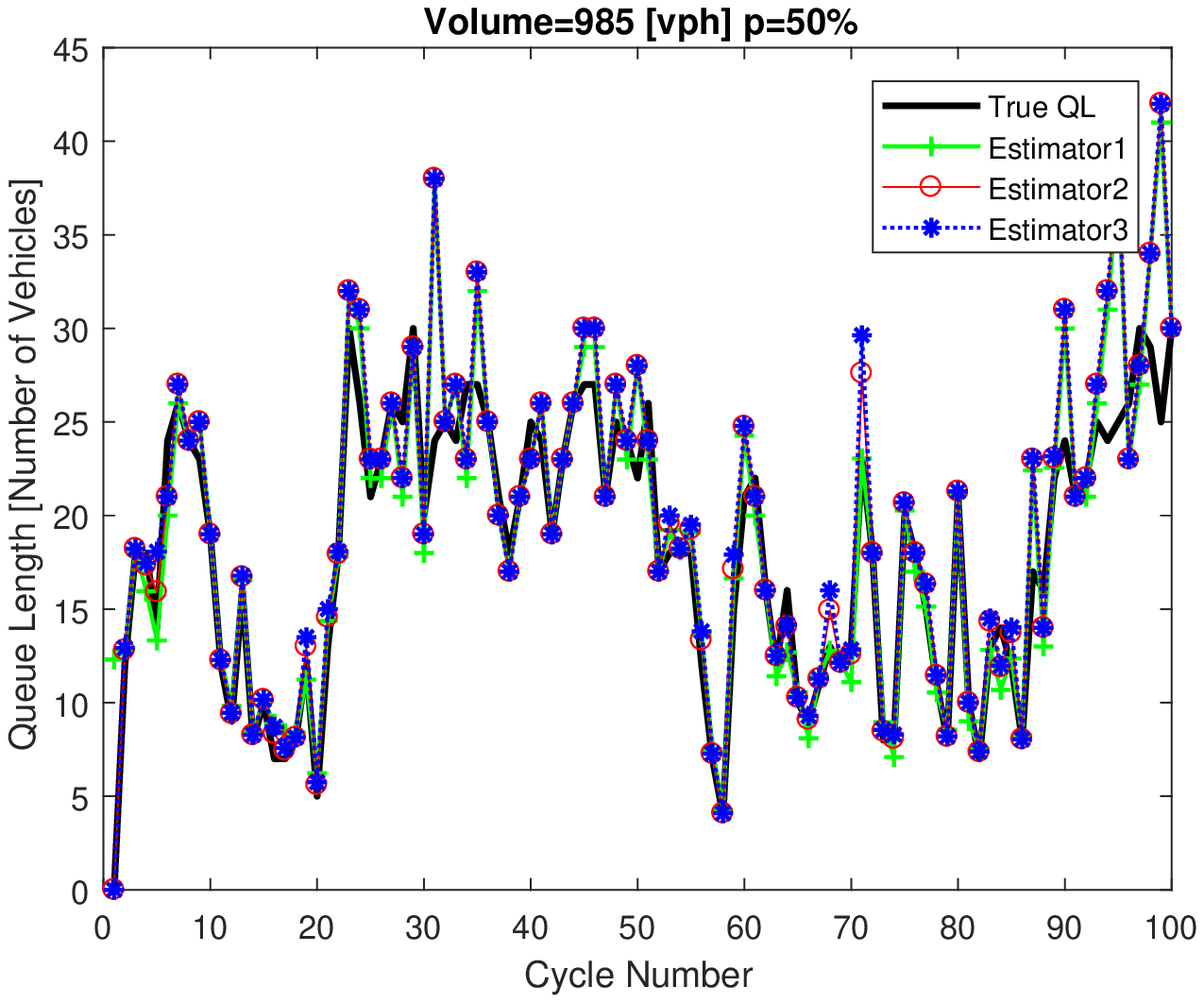}
  \caption{Estimation at $\rho$=0.98}
  \label{fig_10}
\end{subfigure}
\begin{subfigure}{.49\textwidth}
 \centering
\includegraphics[width=1\linewidth]{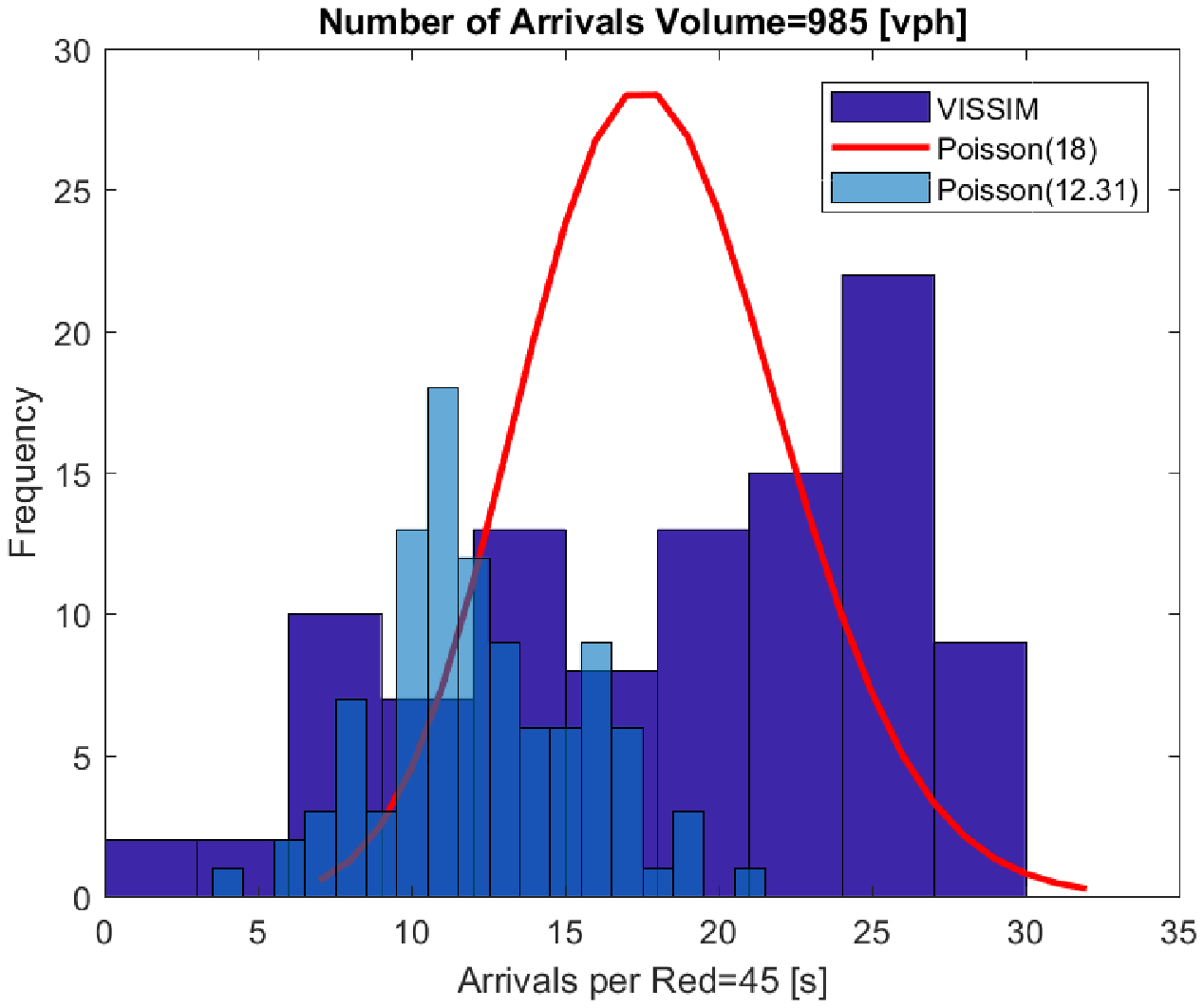}
  \caption{Distribution at $\rho$=0.98}
  \label{fig_10h}
\end{subfigure}
\caption{Cycle-by-cycle QL estimation for $\rho\geq0.80$ values at $p=50 \%$}
\label{fig_high}
\end{figure}

\begin{table}[H]
\centering
\caption{Cycle-to-cycle $\%$ $\sqrt{CV} (RMSE)$ and $\%{EN_i}$ differences of the QL estimators with/without unknown parameters and true QLs at $p=50\%$}
\label{tab_woQ}       % Give a unique label
%\resizebox{\textwidth}{!}{%
\scalebox{0.75}{
\begin{tabular}{l c|c c c |c c c |c c c}
%\hline\noalign{\smallskip}
%& \multicolumn{2}{l}{Interval} \\ 
% & \multicolumn{1}{c}{1-lag} & \multicolumn{1}{c}{5-lag}
%& \multicolumn{1}{c}{10-lag} & \multicolumn{1}{c}{15-lag} & \multicolumn{1}{c}{30-lag} & \multicolumn{1}{c}{45-lag}\\
%Interval 
%\hline\noalign{\smallskip}
% \multicolumn{4}{c}{$\%\Delta{CV_i} \hat{p}_2 \hat{\lambda}_2$} & \multicolumn{3}{c}{$\%\Delta{CV_i} \hat{p}_5 \hat{\lambda}_6$} & \multicolumn{3}{c}{$\%\Delta{EN_i} \hat{p}_2 \hat{\lambda}_2$}& \multicolumn{3}{c}{$\%\Delta{EN_i} \hat{p}_5 \hat{\lambda}_6$} \\

\noalign{\smallskip}\hline\noalign{\smallskip}

  &True QL& Est. 1 &Est. 2 &Est. 3& RMSE 1 & RMSE 2 &RMSE 3& $\%\sqrt{CV_{1}}/{EN_i}$ &\%$\sqrt{CV_{2}}/{EN_i}$& $\%\sqrt{CV_{3}}/{EN_i}$ \\
\noalign{\smallskip}\hline\noalign{\smallskip}

\multirow{1}{*}{$\rho$=0.50} & 5.65 &	5.83 & 5.98 &	6.36 &	1.30 &	0.72	& 1.33 &	23\% &	13\% &	24\% \\
%\hline
\multirow{1}{*}{$\rho$=0.60}	&7.18	&7.08	&7.32	&7.59	&1.43 & 	1.04& 	1.28 &	20\%& 	15\%& 	18\% \\
%\hline
\multirow{1}{*}{$\rho$=0.70}	& 8.39	&8.27	&8.58	& 8.87	&1.73	& 1.02& 	1.19& 	21\%& 	12\%& 	14\% \\
%\hline 
\multirow{1}{*}{$\rho$=0.80}	&10.31	&10.29	& 10.67	& 10.93	&1.56	& 1.39& 	1.67& 	15\%& 	14\%& 	16\% \\
% \hline
\multirow{1}{*}{$\rho$=0.88}	& 12.52	&12.73	&12.99	& 13.14	&2.06	& 1.60& 	1.72& 	16\%& 	13\%& 	14\% \\
%\hline
\multirow{1}{*}{$\rho$=0.985}	& 18.01	& 18.81	&19.15	&19.26	& 3.45	& 3.37& 	3.45& 	19\%& 	19\%& 	19\% \\
\hline
\end{tabular}
}
\end{table}
VISSIM generates vehicles with exponential interarrival times at the origin that move and queue realistically. The arrival profile changes as vehicles move along the network based on vehicle composition, vehicle characteristics, driving behavior, number of lanes, and other network settings. Figs.~\ref{fig_low} and \ref{fig_high} show the behavior of QL estimators and arrivals within red duration when Poisson (random) arrivals are assumed. The figures were obtained for different volume-to-capacity ratios ($\rho$) at $p$=$50\%$. Queue lengths are closely followed by the estimators with $\pm$ $1$ vehicle for $\rho \leq0.88$. Up to $\rho=0.88$, the histograms from VISSIM arrivals are close to the simulated distribution of Poisson random values and therefore Poisson arrival assumption is approximately valid up to $\rho=0.88$.

\subsubsection{Overflow Queue Case: $Q > 0$}
\label{sctqlewq}
In this subsection, real-time estimation of cycle-to-cycle as well as steady-state behavior of QLs when $Q>0$ are presented. The total QL at the end of the red duration for this case is written as sum of queues for three different scenarios as shown in Eq.~(\ref{eqn_ewqi}) below. In this equation, $I(l\in Q)$ is an indicator function for the case of last CV  in overflow queue, $I(l\in A)$ indicates the last CV in new arrivals, and $I(l=0)$ indicates that no CV is present in the queue. 
At any given cycle, the three terms on the right hand side of equation (\ref{eqn_ewqi}) represent disjoint events.  Hence, only one of the three terms will be positive and the remaining two terms will be zero.
\begin{equation}   
E(N_{i}|L=l,T=t,Q_{i}\geq 0)= I(l\in Q)[l+\hat{\theta} (C-t')+\hat{\theta} R]+ I(l\in A)[l+\hat{\theta} \delta]+I(l=0)[(1-\hat{p})(E(Q_{i})+\hat{\theta} R)]
%E(N_{i}|L=l,T=t,Q_{i}\geq 0)=\begin{cases} I(l\in Q)[l+\hat{\theta} (C-t')+\hat{\theta} R]+\\ I(l\in A)[l+\hat{\theta} \delta]+
%\I(l=0)[(1-\hat{p})(E(Q_{i})+\hat{\theta} R)]\end{cases}   
\label{eqn_ewqi}  
\end{equation}

Cycle-by-cycle overflow queue can be given as

\[
 E(Q_{i})=\frac{Xi(\hat{\rho}-1)}{4} \sqrt{(\hat{\rho}-1)^2+\frac{12(\hat{\rho}-\rho_{o})}{Xi}}
\]
where, $\rho_{o}$=$0.67+X/600$, $X=24$ vehicles per cycle, $\hat{\rho}=\hat{\lambda} C/X$ is adopted from \cite{Akcelik1980}.
For all numerical examples in this paper,  $i=1,2,3...$ denotes the cycle index . 
\newline
Another form of cycle-by-cycle overflow queue is given 
\[
 E(Q_{i})=E(Q)(1-e^{-\beta i}) \quad\mbox{where}\quad E(Q)=\frac{3(\hat{\rho}-\rho_{o})}{2(1-\hat{\rho)}}
\]
from~\cite{Viti2006} can also be used and this gives very close results with \cite{comert2013simple}. 
Cycle-to-cycle error of the estimator in Eq.~(\ref{eqn_ewqi}) (i.e., $V(D_{i})$
 with
\[ 
V(Q_{i}) = [E(Q)(\hat{\rho}+(1-\hat{p})/0.15)+(\sqrt{\hat{\rho} Xi}-\sigma_{Q_{e}})e^{-\beta i}]^2
\]
where $\sigma_{Q_{e}}$ can be calculated from 
\[
\sigma_{Q_{e}}=E(Q)(\hat{\rho}+(1-\hat{\rho})/0.15) 
\]
can be given as follows (\cite{Viti2006}),
\begin{equation}   
V(D_{i}|Q_{i}\geq 0)=  P(l\in Q)[\hat{\theta} (C-E(T'))+\hat{\theta} R]+ P(l\in A)[(1-\hat{p})(1-e^{-p\hat{\lambda} R})/\hat{p}]+ P(l=0)[(1-\hat{p})(V(Q_{i})+\hat{\theta} R)]
\label{eqn_vdwqi}
\end{equation}
%V(D_{i}|Q_{i}\geq 0)=\begin{cases} P(l\in Q)[\hat{\theta} (C-E(T'))+\hat{\theta} R]+\\ P(l\in A)[(1-\hat{p})(1-e^{-p\hat{\lambda} %R})/\hat{p}]+\\P(l=0)[(1-\hat{p})(V(Q_{i})+\hat{\theta} R)]\end{cases}   
%\label{eqn_vdwqi}  
%\end{equation}

Table~\ref{tab_summary} summarizes the queue length estimation errors when overflow queue is included. The errors are expressed in $\%\Delta{CV_i}$ and $\%\Delta{EN_i}$ where, $CV_{i}$=$\sqrt{V(D_{i})}/E(N_{i})$. Since estimators are able to capture overflow queue, errors are declining to zero as CV  penetration rate increases. Results from literature reviewed confirm that 
$p\geq30\%$ queue length can be predicted within $\pm10\%$ across all all $\rho$ levels. At $p=50\%$, queue length can be predicted within $\pm2\%$ in $\%\Delta{CV_i}$ and $\%\Delta{EN_i}$. In section~\ref{sct_numex}, we present numerical results for queue length-based signal control  when queue length estimation errors are ignored with $p\geq50\%$. Thus,we can use  any of the estimators (estimator 2 or estimator 3)  in Eqs. (\ref{eqn_gmax1}) and (\ref{eqn_gmax2}).

\begin{table}[h!]
\centering
\caption{$i$=$2$ cycle-to-cycle $\%$ $CV_{i}$ and $\%{EN_i}$ differences of the QL estimators with unknown parameters and true QLs}
\label{tab_summary}       % Give a unique label
%\resizebox{\textwidth}{!}{%
\scalebox{0.7}{
\begin{tabular}{l |c c c |c c c |c c c |c c c}
%\hline\noalign{\smallskip}
%& \multicolumn{2}{l}{Interval} \\ 
% & \multicolumn{1}{c}{1-lag} & \multicolumn{1}{c}{5-lag}
%& \multicolumn{1}{c}{10-lag} & \multicolumn{1}{c}{15-lag} & \multicolumn{1}{c}{30-lag} & \multicolumn{1}{c}{45-lag}\\
%Interval 
%\hline\noalign{\smallskip}
 \multicolumn{4}{c}{$\%\Delta{CV_i} \hat{p}_1 \hat{\lambda}_1$} & \multicolumn{3}{c}{$\%\Delta{CV_i} \hat{p}_2 \hat{\lambda}_2$} & \multicolumn{3}{c}{$\%\Delta{EN_i} \hat{p}_1 \hat{\lambda}_1$}& \multicolumn{3}{c}{$\%\Delta{EN_i} \hat{p}_2 \hat{\lambda}_2$} \\

\noalign{\smallskip}\hline\noalign{\smallskip}

 probe &$\rho=0.985$& $\rho=0.88$ &$\rho=0.70$ &$\rho=0.985$& $\rho=0.88$ & $\rho=0.70$ &$\rho=0.985$& $\rho=0.88$ &$\rho=0.70$ &$\rho=0.985$& $\rho=0.88$ & $\rho=0.70$ \\
\noalign{\smallskip}\hline\noalign{\smallskip}

\multirow{1}{*}{p=0.1\%} & 1\% &	-1\% & 0\% &	1\% &	-1\% &	0\%	& 99\% &	100\% &	98\% &	98\% &	95\% &	98\% \\
%\hline
\multirow{1}{*}{p=5\%}	&-6\%	&-6\%	&-12\%	&-5\%	&-5\% & 	-15\%& 	70\% &	83\%& 	80\%& 	63\%& 	74\%& 	68\% \\
%\hline
\multirow{1}{*}{p=10\%}	& 7\%	&-9\%	&-8\%	& 6\%	&-8\%	& -13\%& 	55\%& 	49\%& 	62\%& 	48\%& 	28\%& 	44\%\\
%\hline 
\multirow{1}{*}{p=20\%}	&-1\%	&-3\%	& 3\%	& 0\%	&-1\%	& 0\%& 	24\%& 	8\%& 	12\%& 	13\%& 	0\%& 	4\%\\
% \hline
\multirow{1}{*}{p=30\%}	& 1\%	&-7\%	&-7\%	& 0\%	&-6\%	& -8\%& 	6\%& 	6\%& 	-1\%& 	-2\%& 	3\%& 	-5\%\\
%\hline
\multirow{1}{*}{p=50\%}	& 0\%	& 0\%	&-1\%	&-1\%	& 0\%	& -1\%& 	1\%& 	2\%& 	2\%& 	0\%& 	0\%& 	0\%\\
\hline
\end{tabular}
}
\end{table}
\begin{eqnarray}
E(N|l,t,m,Q_{i}=0)=l+(l-m/l)(l/R)(R-t)=l+(l-m)(1-t/R)
\label{eqn_eltapprox2a} 
\end{eqnarray}
\begin{eqnarray}
E(N|l,t,m,Q_{i}=0)=l+(1-\frac{mt}{mt+(l-m)R})((l-m)/t+m/R)(R-t)=m+\frac{R(l-m)}{t}
\label{eqn_eltapprox9a} 
\end{eqnarray}
\section{Numerical Experiments}
\label{sct_numex}
\begin{figure}[ht!]
\centering
\includegraphics[scale=.35]{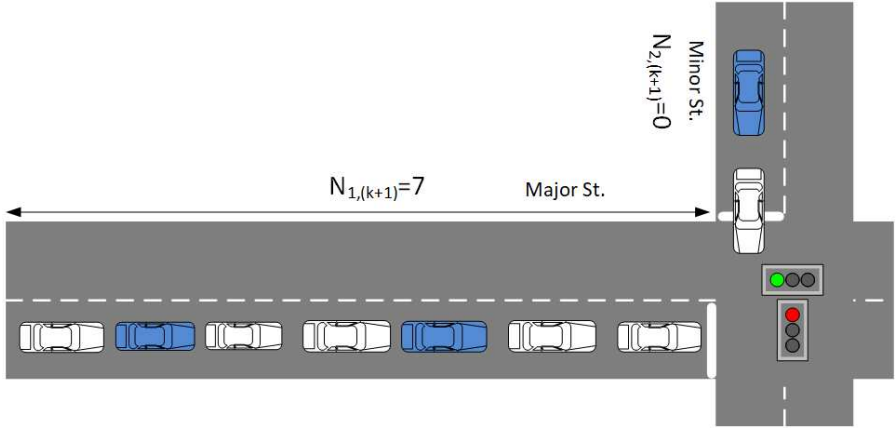}
\caption{Intersection used in the experiments}
\label{fig_int}       
\end{figure}

Numerical experiments are conducted at randomly selected demand levels to investigate how $\beta$, $LB_1$, and $LB_2$ interact and to evaluate the benefits of using $100 \%$ accurate QL information (i.e., at $p=100 \%$ CV market penetration level) in the proposed signal control as compared to the typical actuated control. A simple intersection where a one-way major road intersects with a minor street as shown in Fig. \ref{fig_int}, is designed in  VISSIM. Two signal control methods are designed and evaluated. A typical actuated signal control is utilized with fixed max green. The proposed queue-length based method (i.e., QL-based) is given with max green times that are determined based on Eqs. (\ref{eqn_gmax1}) and (\ref{eqn_gmax2}). It is assumed that there are stop bar loops on both approaches (i.e., fully actuated control). Real-time data from these loops are used in both signal control methods to extend the greens. 

Parameters are chosen as described in section \ref{sct_parsel} for three different base demand profiles for both signal control methods over one-hour period. Table \ref{tab_demand} shows these demand profiles for each approach. These demand scenarios and the respective optimum signal control parameters for them are adopted from \cite{Comert2009TRB}. For profile 1,  the demand exhibits a steep increase in the second quarter and a decrease in the last quarter for major and minor streets. Demand profile 2 has rather high demand and a large disparity between major and minor streets. Demand profile 3 has larger jumps in demands than the other two profiles. 

\begin{table}[!ht]
\centering
\caption{Three base demand profiles}
\label{tab_demand}
\scalebox{0.8}{     
\begin{tabular}{c c c c c c c}
%p{1.5cm}p{2cm}p{2cm}p{2.4cm}p{2cm}
\hline\noalign{\smallskip}
& \multicolumn{2}{c}{Profile\#1} & \multicolumn{2}{c}{Profile\#2} & \multicolumn{2}{c}{Profile\#3} \\
Time(min)&Major St.& Minor St.& Major St.& Minor St.& Major St.& Minor St.\\
\noalign{\smallskip}\hline\noalign{\smallskip}
$0-15$ & $800$ & $300$ & $1200$ & $100$ & $800$ & $300$\\
$16-30$ & $1350$ & $400$ & $1400$ & $200$ & $1500$ & $400$\\
$31-45$ & $1350$ & $400$ & $1500$ & $300$ & $900$ & $200$\\
$46-60$ & $800$ & $300$ & $1100$ & $100$ & $500$ & $100$\\
\noalign{\smallskip}\hline\noalign{\smallskip}
\end{tabular}
}
\end{table}
The volumes in Table \ref{tab_demand} are assigned to $15$-minute intervals for a total of one hour simulations. These base volumes are used to determine the parameters of both typical actuated signal (i.e., $gmax1$, $gmax2$) and QL-based adaptive actuated signal (i.e., $\beta$, $LB1$, and $LB2$). These optimal parameters are found based on the simulation of random arrivals on both major and minor streets (i.e., no upstream signal on the major street). Other parameters of the isolated actuated signal are set to be identical for both control methods (i.e., $gmin1$=$10$ s, $gmin2$=$5$ s, $all red$=$1$ s, $yellow$=$2$ s, $gap out$=$3$ s). The same random seed is used in simulating both methods in VISSIM to ensure that the same vehicle patterns are generated. The optimal  parameter values for all the three demand profiles are shown in Table \ref{tab_par}. 
It is observed that there isn't a very significant difference between $gmax_1$ and $LB_1$ and also between $gmax_2$ and $LB_2$. The $\beta$ values are the same for these demand profiles.

\begin{table}[!ht]
\centering
\caption{Optimum parameters for the three demand profiles}
\label{tab_par}
\scalebox{0.8}{     
\begin{tabular}{c|c c|c c c}
%p{1.5cm}p{2cm}p{2cm}p{2.4cm}p{2cm}
\hline\noalign{\smallskip}
& \multicolumn{2}{c}{Typical} & \multicolumn{2}{c}{QL-Based} \\
Demand Profile&$gmax_1$(s) & $gmax_2$(s) & $\beta$ & $LB_1$(s)& $LB_2$(s) \\
\noalign{\smallskip}\hline\noalign{\smallskip}
$1$ & $75$ & $15$ & $2.5$ & $55$ & $10$ \\
$2$ & $70$ & $20$ & $2.5$ & $75$ & $10$ \\
$3$ & $60$ & $20$ & $2.5$ & $55$ & $10$ \\
\noalign{\smallskip}\hline\noalign{\smallskip}
\end{tabular}
}
\end{table} 
\vspace{-15pt}
\section{Sensitivity Analysis}
\label{sct_sens}
In this section, we compare the robustness of the QL-based and typical actuated methods to demand fluctuations  by changing the base volumes on the minor and major streets by $\pm 20\%$ but keeping the corresponding optimal parameters shown in Table \ref{tab_par} constant. 
%In addition, to evaluate the robustness to vehicle arrival patterns,
 Two arrival types are considered: (i) random and (ii) platoon arrivals where an upstream signal with sufficiently large capacity is introduced in the upstream of the major intersection. This signal has a fixed cycle length of $92$ seconds and $60$ seconds phase lengths for the major street. The major road has two lanes at the upstream signal location whereas it narrows to one lane between the upstream and subject intersection. It is designed this way to ensure that the upstream signal does not form a bottleneck but only serves to generate platoon arrivals. 

\begin{table*}[!ht]
\centering
\caption{Performance measures for each demand scenario for random arrivals}
\label{tab_random}
%\resizebox{\linewidth}{!}{
\scalebox{0.75}{
\begin{tabular}{l l l c c c c c c c c c}
\hline\noalign{\smallskip}
& & & \multicolumn{3}{c}{AvgDelay(s)} & \multicolumn{3}{c}{NStops} & \multicolumn{3}{c}{AvgQueue(m)} \\
\cline{2-12}
& & & Typ&QL-B&\%Imp&Typ&QL-B&\%Imp&Typ&QL-B&\%Imp\\
\noalign{\smallskip}\hline\noalign{\smallskip}
\multirow{4}{*}{Profile1}&1-1&+20\%+20\% & 39.94 & 21.98 & 45\% & 3.84 & 1.80 & 53\% & 66.96 & 20.76 & 69\%\\
&1-2&+20\%-20\%& 15.88 & 15.59 & 2\% & 1.11 & 1.12 & -1\% & 12.59 & 10.94 & 13\%\\
&1-3&-20\%+20\%& 12.20 & 1.17 & 0\% & 0.75 & 0.75 & 0\% & 8.85 & 8.76 & 1\%\\
&1-4&-20\%-20\%& 9.80 & 9.45 & 4\% & 0.62 & 0.59 & 5\% & 5.47 & 5.22 &  5\%\\
\cline{2-12}
\multirow{4}{*}{Profile2}& 2-1 & +20\%+20\% & 15.34 & 15.45 & -1\% & 1.21 & 1.12 & 7\% & 12.04 & 12.82 & -6\%\\
&2-2&+20\%-20\%& 12.18 & 11.52 & 5\% & 0.88 & 0.79 & 10\% & 7.68 & 7.35 & 4\%\\
&2-3&-20\%+20\%& 9.45 & 8.52 & 10\% & 0.61 & 0.55 & 10\% & 5.71 & 4.80 & 16\%\\
&2-4&-20\%-20\%& 7.24 & 7.07 & 2\% & 0.47 & 0.45 & 4\% & 3.66 & 3.46 &  5\%\\
\cline{2-12}
\multirow{4}{*}{Profile3}& 3-1& +20\%+20\% & 16.48 & 16.43 & 0\% & 1.27 & 1.27 & 0\% & 11.42 & 11.88 & -4\%\\
&3-2&+20\%-20\%& 12.83 & 12.32 & 4\% & 1.00 & 0.89 & 11\% & 7.23 & 6.73 & 7\%\\
&3-3&-20\%+20\%& 11.14 & 10.95 & 2\% & 0.71 & 0.70 & 1\% & 6.50 & 6.40 & 2\%\\
&3-4&-20\%-20\%& 8.74 & 8.55 & 2\% & 0.58 & 0.56 & 3\% & 4.03& 3.83 & 5\%\\
\noalign{\smallskip}\hline\noalign{\smallskip}
\end{tabular}
%}
}
\end{table*}
The results for the random arrivals and platoon arrivals are presented in Table \ref{tab_random} and Table \ref{tab_platoon} respectively. It should be noted that the optimal parameters given in Table \ref{tab_par} for both the actuated and  proposed method are determined under the assumption of random arrival distributions. These optimal parameters are kept the same in all simulation runs.  

For each of the three demand profiles, four different scenarios are modeled and run based on the $\pm 20\%$ demand fluctuations as indicated in the first columns of Table \ref{tab_random} and Table \ref{tab_platoon}. For example, in scenario 1-2 (i.e., demand profile 1 in scenario 2), the base volume of demand profile 1 for the major street is increased by $20\%$ while the minor street volume is decreased by $20\%$. Each scenario is run $30$ times with different random seeds. Average performance measures of these runs are then determined. VISSIM can give output of several performance measures for every desired time interval. From these performance measures, average delay in seconds (s), number of stops, and average queue in meters (m) are obtained at the end of each one-hour run. As the results below indicate, overall the QL-based method  perform better than the typical actuated control at all demand profiles, scenarios, and performance measures.

\begin{table*}[!ht]
\centering
\caption{Performance measures for each demand scenario for platoon arrivals}
\label{tab_platoon}       % Give a unique label
%\resizebox{\linewidth}{!}{
\scalebox{0.75}{
\begin{tabular}{l l l c c c c c c c c c}
\hline\noalign{\smallskip}
& & & \multicolumn{3}{c}{AvgDelay(s)} & \multicolumn{3}{c}{NStops} & \multicolumn{3}{c}{AvgQueue(m)} \\
\cline{2-12}
& & & Typ & QL-B & \%Imp & Typ & QL-B & \%Imp & Typ & QL-B & \%Imp\\
\noalign{\smallskip}\hline\noalign{\smallskip}
\multirow{4}{*}{Profile1}& 1-1 & +20\%+20\% & 39.18 & 32.07 & 18\% & 4.89 & 4.09 & 17\% & 25.54 & 13.14 & 49\%\\
& 1-2& +20\%-20\% & 29.83 & 29.17 & 2\% & 4.14 & 4.13 & 0\% & 7.71 & 6.80 & 12\%\\
& 1-3& -20\%+20\% & 14.38 & 13.90 & 3\% & 1.04 & 0.99 & 5\% & 6.43 & 5.97 & 7\%\\
& 1-4& -20\%-20\% & 12.90 & 12.71 & 1\% & 0.97 & 0.94 & 3\% & 3.89 & 3.69 &  5\%\\
\cline{2-12}
\multirow{4}{*}{Profile2}& 2-1 & +20\%+20\% & 37.38 & 36.94 & 1\% & 5.98 & 5.98 & 0\% & 7.02 & 6.46 & 8\%\\
& 2-2& +20\%-20\% & 36.35 & 36.05 & 1\% & 6.16 & 6.14 & 0\% & 4.04 & 3.79 & 6\%\\
& 2-3& -20\%+20\% & 14.55 & 14.26 & 2\% & 1.18 & 1.16 & 2\% & 3.95 & 3.69 & 7\%\\
& 2-4& -20\%-20\% & 13.44 & 13.30 & 1\% & 1.13 & 1.12 & 1\% & 2.55 & 2.44 &  4\%\\
\cline{2-12}
\multirow{4}{*}{Profile3}& 3-1 & +20\%+20\% & 27.09 & 26.80 & 1\% & 3.55 & 3.55 & 0\% & 7.99 & 7.55 & 5\%\\
& 3-2& +20\%-20\% & 25.53 & 25.04 & 2\% & 3.66 & 3.62 & 1\% & 4.49 & 4.07 & 9\%\\
& 3-3& -20\%+20\% & 13.23 & 12.95 & 2\% & 1.01 & 0.99 & 2\% & 4.41 & 4.16 & 6\%\\
& 3-4& -20\%-20\% & 12.06 & 11.73 & 3\% & 0.97 & 0.94 & 3\% & 2.71 & 2.48 & 9\%\\
\noalign{\smallskip}\hline\noalign{\smallskip}
\end{tabular}
%}
}
\end{table*}
From Scenario 1 of demand profile 1 in Table 7, where the volumes are increased by $20\%$ for each direction, 
we see that the improvements reach up to $69\%$ in average QL, $53\%$ in average delay and, $45\%$ in number of stops 

%In Table \ref{tab_random}, the improvements reach up to $69\%$ in average QL, $53\%$ in average delay and, $45\%$ in number %of stops. Scenario 1 of demand profile 1, where the volumes are increased by $20\%$ for each direction, shows these exceptional %results.

\begin{figure}[ht!]
\centering
\includegraphics[scale=.35]{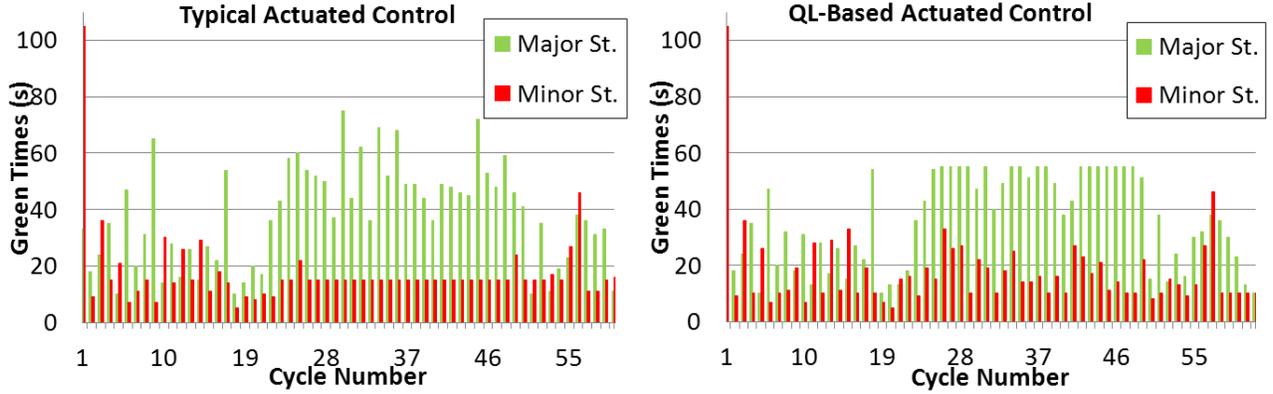}
\caption{Green times and cycle lengths for scenario 1-1 for platoon arrivals}
\label{fig_green11}       
\end{figure}

Compared to profiles 2 and 3, the traffic demand stays at the peak level for $30$ minutes in profile 1 as opposed to $15$ minutes in the other two profiles. Therefore, increasing demand on both major and minor streets by $20\%$ pushes the system to operate at oversaturation level for $30$ minutes. For this scenario the signal is maxing out mostly for the minor street as can be observed in Fig. \ref{fig_green11}. The improvements for platoon arrivals (on the major street) in Table \ref{tab_platoon} reach up to $49\%$ in average QL, $18\%$ in average delay and, $17\%$ in number stops. Scenario 1 of demand profile 1, where the volumes are increased by $20\%$ for each direction, shows these exceptional results. The results in both tables show that the QL-based method performs better for both types of arrivals.

Compared to random arrivals, when the arrivals on the major street are random, the relative improvements are somewhat smaller.
This is expected as the control parameters are optimized based on the random arrivals. It is observed that the number of max-outs decrease significantly for the major street  when arrivals are platoon (e.g., on average $1$ max-out for the one-hour simulation period for platoon arrivals, $17$ for random arrivals for scenario 1-1). On the other hand, number of max-outs remains similar on the minor street (e.g., $33$ for platoon, $30$ for random arrivals for scenario 1-1). Overall, both the typical actuated and QL-based methods perform similar on the major street when the arrivals are platoon. 

In order to see how the two signal control methods compare in terms of performance metrics and allocating the capacity, green distributions and performance measures over the one-hour simulation time are plotted for two selected scenarios: Scenarios 1-1 and 3-1. These two scenarios are chosen to analyze the differences between the typical actuated signal operations and the QL-based operations when the QL-based method performs significantly better (as in Scenario 1-1) and when they perform comparably (as in Scenario 3-1). These comparisons are presented only for the case when arrivals on the major street are platoon.  The results for the random arrivals are very similar. 

\begin{figure}[ht!]
\centering
\includegraphics[scale=.30]{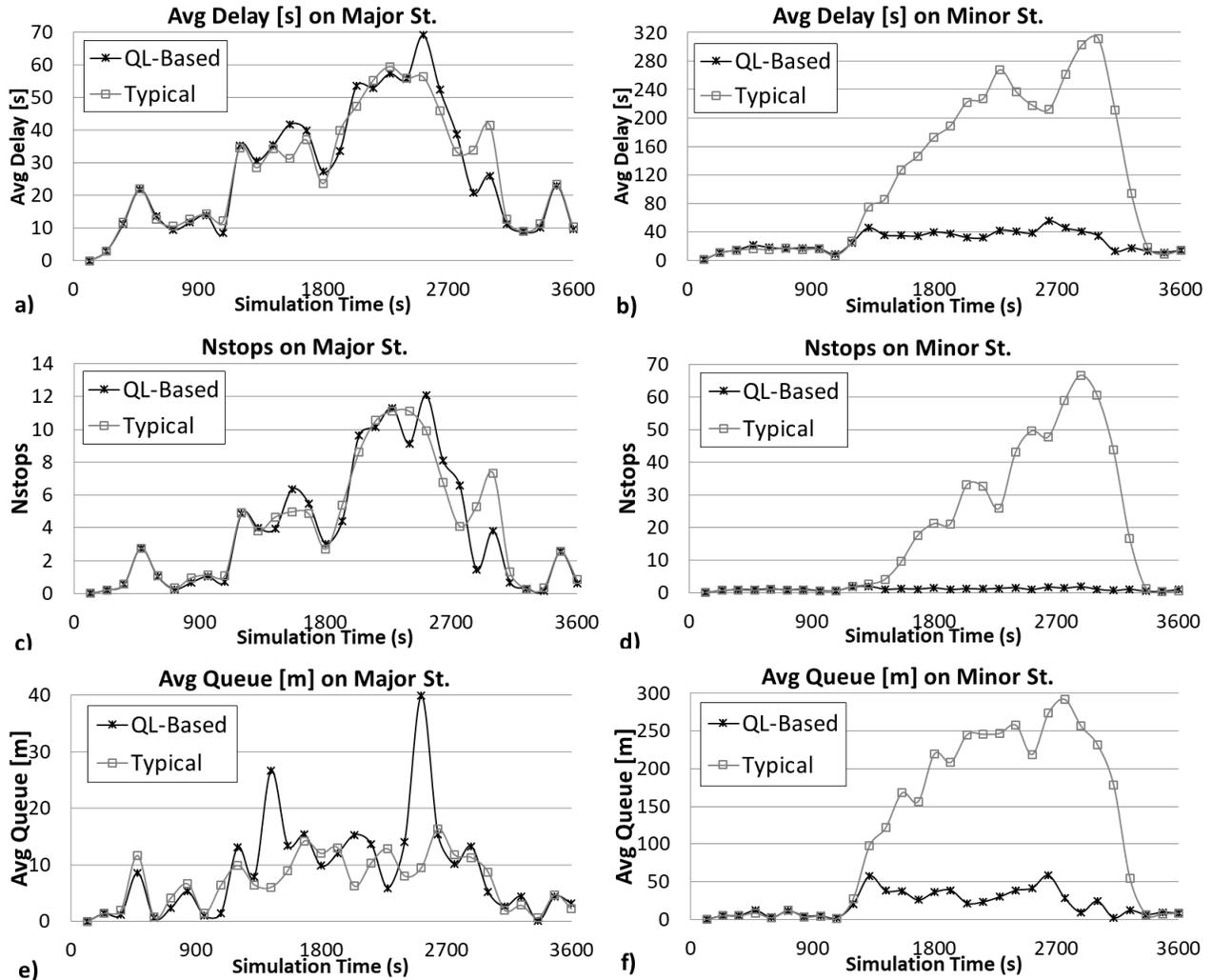}
\caption{Performance measures for both major and minor streets obtained with the typical actuated control and QL-based control for scenario 1-1 for platoon arrivals}
\label{fig_scn11}       
\end{figure}

Figs. \ref{fig_scn11}-a and \ref{fig_scn11}-b show the variation in average delay on major and minor streets, respectively, for both control methods. Clearly, the QL-based method produces much lower delays than the typical actuated method, particularly on the Minor Street. Likewise, Figs. \ref{fig_scn11}-c to \ref{fig_scn11}-f show the variation in average QL and number of stops. It can be seen that the queue on the minor street grows dramatically under the typical actuated control. The average delay and QL on the minor street increase dramatically since the max green under the typical actuated control for the minor street (i.e., $15$ s) is not sufficient to deal with the increase in demand. 

\begin{figure}[ht!]
\centering
\includegraphics[scale=.30]{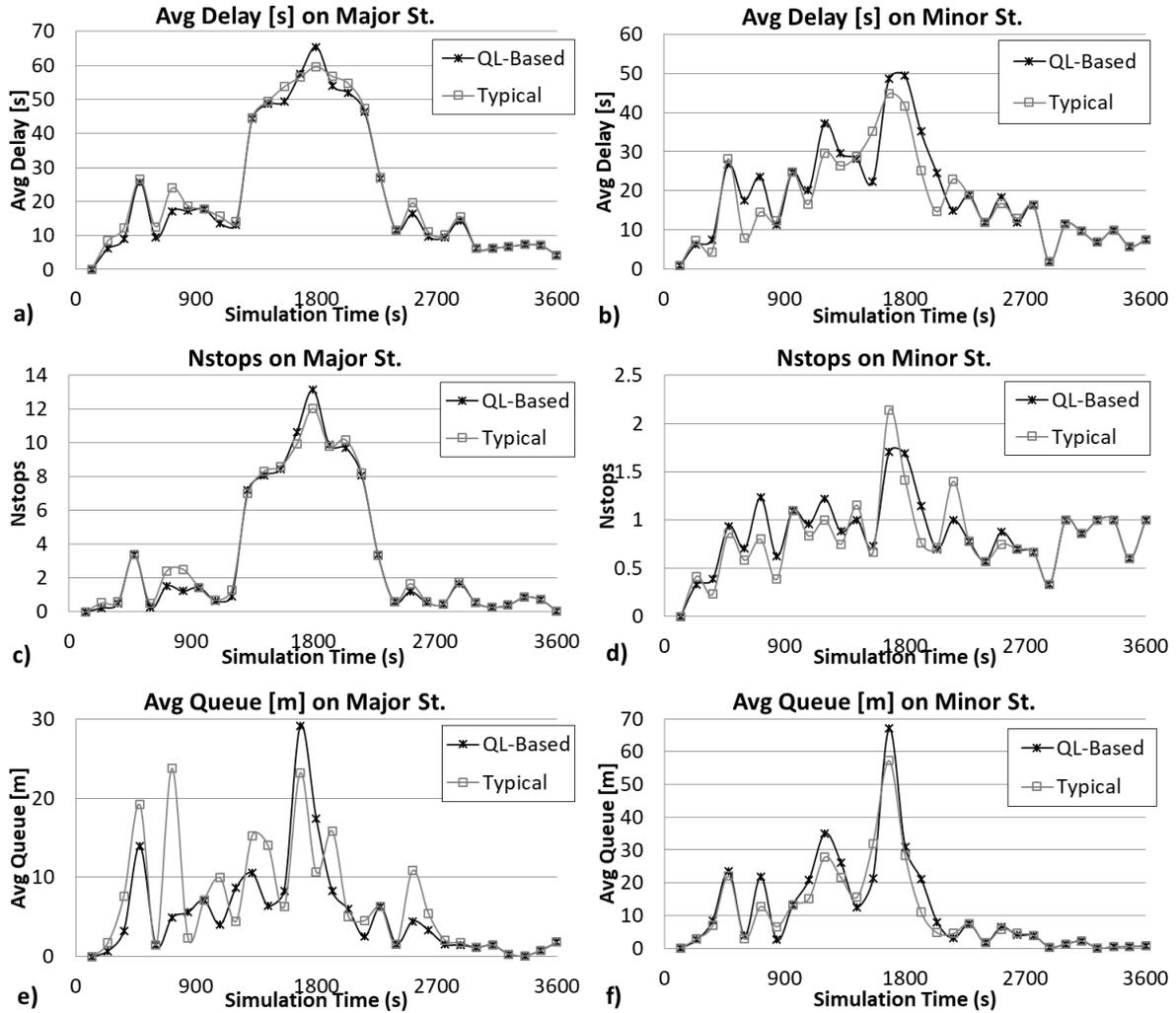}
\caption{Performance measures for both major and minor streets obtained from the typical actuated control and QL-based control for scenario 3-1 for platoon arrivals}
\label{fig_scn31}       
\end{figure}

Fig. \ref{fig_scn31} shows the scenario with almost identical performance measures for both  types of control methods. Under scenario 1 of demand profile 3, minor street traffic stays manageable with $20$ seconds max green (e.g., there are $11$ max outs for scenario 3-1 as opposed to $30$ for scenario 1-1). Hence, in this case, the QL-based method performs very similarly since the average QL does not grow substantially to induce significant changes. This can be seen from Fig. \ref{fig_scn31}-f. In the typical actuated control, the average QL is about $65$ m for scenario 3-1 and  $300$ m for scenarios 1-1. 

\begin{figure}[ht!]
\centering
\includegraphics[scale=.35]{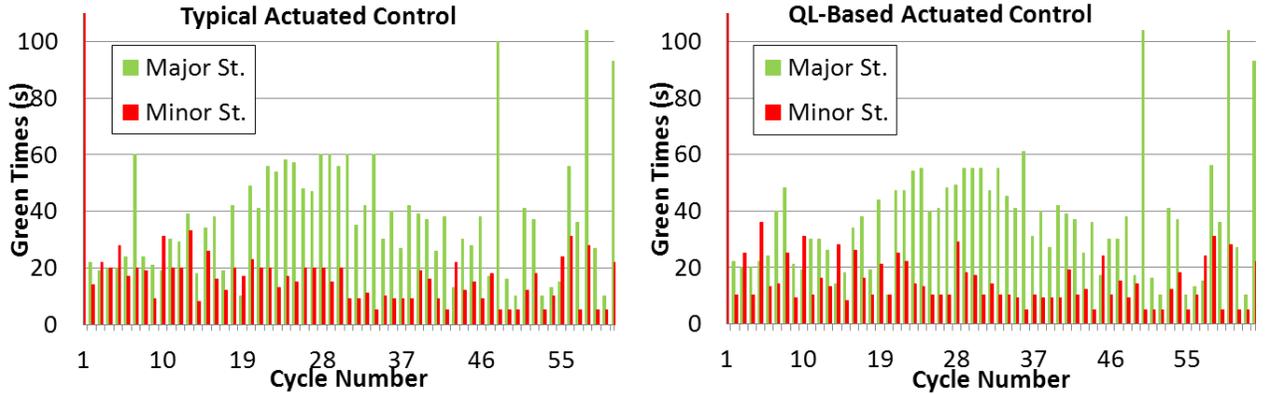}
\caption{Green times and cycle lengths for scenario 3-1 for platoon arrivals}
\label{fig_green31}       
\end{figure}

Figs. \ref{fig_green11} and \ref{fig_green31} show the actual green times by cycle number for both control methods. Under the typical actuated control, the minor street is maxing out many times, especially between cycles $26$ and $50$ when the volumes are larger. Note that at some cycles the green time of the minor street exceeds the max green of $15$ seconds. This happens since the max green timer counts down when there is call on the opposing phase. The QL-based method changes the max green between lower and upper bound. For the minor street, the green times are larger than the lower bound ($10$ seconds) in a number of cycles to accommodate the large volumes.

Fig. \ref{fig_green11} shows the total green times for both methods for scenario 1-1. It can be observed that the cycle lengths in the QL-based method and the typical actuated control are quite similar. However, the green distributions to major and minor streets differ significantly as shown in Fig. \ref{fig_green11}. The QL-based method takes away the green from the major street to allocate more time to the minor street to prevent the queue to grow substantially.
% as it does in the typical actuated control case.
 It is clear that the proposed method is effective in accommodating imbalances in the queues on the two approaches as illustrated in Fig. \ref{fig_scn11} for this scenario. In conclusion, the QL-based method adjusts the greens more efficiently to adapt to changes in demand for the two conflicting signal phases.  

Fig. \ref{fig_green31} demonstrates the green times and the cycle times for scenario 3-1.
% As shown in Fig. \ref{fig_green31} below, 
The performance of both control types are quite similar for this particular scenario. Consequently, the major and minor street green distributions produced by the two control methods are also similar.
\begin{table}[!ht]
\centering
\caption{Average improvements over the typical actuated control}
\label{tab_all} 
\scalebox{0.75}{    
\begin{tabular}{l c c c |c c c}
%p{1.5cm}p{2cm}p{2cm}p{2.4cm}p{2cm}
\hline\noalign{\smallskip}
& \multicolumn{3}{c}{Random Arrivals} & \multicolumn{3}{c}{Platoon Arrivals} \\
\cline{2-7}
& Overall & Major St. & Minor St. & Overall & Major St. & Minor St. \\
\noalign{\smallskip}\hline\noalign{\smallskip}
AvgDelay & $6\%$ & $3\%$ & $7\%$ & $3\%$ & $1\%$ & $5\%$ \\
NStops & $9\%$ & $6\%$ & $3\%$ & $3\%$ & $2\%$ & $3\%$ \\
AvgQueue & $10\%$ & $6\%$ & $8\%$ & $11\%$ & $9\%$ & $7\%$ \\
\noalign{\smallskip}\hline\noalign{\smallskip}
\end{tabular}
}
\end{table}

The overall results for all the 12 scenarios are summarized in Table \ref{tab_all} for both random and platoon arrivals. Since each scenario is run 30 times and there are a total of 4 cases to be considered (two control types and two arrival types), the total number of simulation runs for the complete analysis is $1,440$. This table gives an aggregate summary of all these runs and shows the average percent improvements in average delay, number of stops, and average QL that are achieved when the QL-based method is compared to the typical actuated control for all scenarios. The breakdown of the improvements by major and minor streets is also shown. Based on the results, the delay savings are mostly associated with the minor street.
\vspace{-15pt}
\section{Conclusions}
This paper presents a new queue length based control method for signals where max green times are calculated in each cycle based on the measured QLs. By varying max green times from cycle to cycle based on a simple formula, the intersection capacity is allocated more efficiently when traffic demand fluctuates. The proposed method is implemented for a single intersection with random and platoon vehicle arrivals and its performance is evaluated in a microscopic traffic simulation environment (i.e., VISSIM). To assess the robustness of the proposed method, various numerical experiments are conducted where traffic demand on the intersection approaches is increased and decreased by $20\%$ relative to the demand levels for which signal timing parameters are optimized. Compared to the typical fully-actuated signal operations, the proposed queue-length-based method provides significant improvements in efficiency in terms of average delay, number of stops, and queue size. The results show significant potential benefits of using QL information. Overall, when all scenarios are considered, the average delay of the isolated intersection was improved by $3\%$, number of stops by $3\%$, and average QL by $11\%$. For individual scenarios, much larger improvements ranging from 50 to 60\% are observed. Future research is needed to analyze more complicated intersections with more than two phases.
\vspace{-15pt}
\section*{Acknowledgments}
This study is supported by the Center for Connected Multimodal Mobility ($C^2$$M^2$) (USDOT Tier 1 University Transportation Center) Grant headquartered at Clemson University, Clemson, South Carolina, USA. The authors would also like to acknowledge U.S. Department of Homeland Security (DHS) Summer Research Team Program Follow-On, U.S. Department of Education Minority Science Improvement Program (MSIEP-Benedict College Creating Achievers in STEM through Academic Support, Mentoring, and Research), and U.S. National Science Foundation (NSF, Nos. 1719501, 1436222, and 1400991) grants. 

Any opinions, findings, conclusions or recommendations expressed in this paper are those of the author(s) and do not necessarily reflect the views of ($C^2$$M^2$), U.S. DOT, U.S. DHS, U.S. Department of Education, or NSF and the U.S. Government assumes no liability for the contents or use thereof.

\bibliographystyle{model5-names}
\bibliography{transport_bibliography}

\end{document}